\documentclass[12pt,preprint]{aastex}

\slugcomment{accepted by ApJ 29 Sept. 2005, scheduled for Feb. 2006 issue}

\newcommand{\etal}{et\,al.}

\begin{document}

\title{Pixie Dust: The Silicate Features in the Diffuse Interstellar Medium}

\author{J.E. Chiar\altaffilmark{1} and A.G.G.M. Tielens\altaffilmark{2}}

\altaffiltext{1}{SETI Institute, 515 N. Whisman Drive, Mountain View,CA
94043 and NASA-Ames Research Center, Mail Stop 245-3, Moffett
Field, CA 94035} 

\altaffiltext{2}{SRON/Kapteyn Astronomical Institute, P.O. Box 800, 9700
AV Groningen, The Netherlands}

\begin{abstract}
We have analyzed the 9.7 and ``18'' \micron\ interstellar silicate absorption
features along the line of sight toward four heavily extincted galactic WC-type
Wolf-Rayet (WR) stars. We construct two interstellar extinction curves from 1.25
to 25 \micron\ using near-IR extinction measurements from the literature along
with the silicate profiles of WR 98a (representing the local ISM) and GCS 3
(representing the Galactic Center).  We have investigated the mineralogy of the
interstellar silicates by comparing extinction profiles for amorphous silicates
with olivine and pyroxene stochiometry to the 9.7 and ``18'' \micron\ absorption
features in the WR 98a spectrum. In this analysis, we have considered solid and
porous spheres and a continuous distribution of ellipsoids. While it is not
possible to simultaneously provide a perfect match  to both profiles, we find the
best match requires a mixture of these two types of compounds. We also consider
iron oxides, aluminosilicates and silicate carbide (SiC) as grain components.
Iron oxides cannot be accommodated in the observed spectrum, while the amount of
Si in SiC is limited to $<4$\%.  Finally, we discuss the cosmic elemental
abundance constraints on the silicate mineralogy, grain shape and porosity.    

\end{abstract}

\keywords{ISM: lines and bands --- ISM: abundances --- ISM: dust ---
ISM: extinction --- ISM: molecules --- infrared: ISM}

\section{Introduction}

Astronomical observations are often hampered by the confusing presence of
foreground dust. These data are then adjusted using vaguely understood
correction procedures, which are perhaps best described as  ``faith, trust, and
pixie dust.'' In this, astronomers follow notions which are  widely spread in
society because the term ``pixie dust'' is often used to refer to a technology
that seemingly does the impossible. In this paper, we address one aspect of
interstellar pixie dust: the nature of interstellar silicates and the infrared
extinction ``correction'' procedure.

The presence of silicate dust in the  interstellar medium has been known for
some four decades, but its precise mineralogy remains uncertain. Silicates in
the interstellar medium (ISM) are thought to be the product of mass-losing
oxygen-rich Asymptotic Giant Branch (AGB) stars.  Amorphous silicates are found
in the circumstellar environments of almost all evolved oxygen-rich mass-losing
stars \nocite{kemper02thesis} (Kemper 2002).  Crystalline silicates are found
around AGB stars with high mass-loss rates, although it's possible these
silicates exist but are undetectable around low mass-loss AGB stars (Kemper
2002).  In the ISM,  crystalline silicates are not detected and, recently, a
stringent limit of $<1$\% of the total silicate population by mass has been
placed\nocite{kemper_vriend_tielens04,kemper_vriend_tielens05} (Kemper, Vriend,
\& Tielens 2004, 2005).  

Amorphous silicates in the ISM exhibit absorption features with smooth profiles
at 9.7 and $\sim18$ \micron.  These profiles can be fitted with extinction
features based on laboratory derived optical constants.  This procedure results
in estimates for abundances in solid form of Si, Mg and Fe, which depend on the
assumed porosity, particle shape, and the adopted silicate material (olivine or
pyroxene-glass, \nocite{mathis98} Mathis 1998). Typically, such estimated require most
or all of the available solar abundance of Si, Mg, and Fe.

Bright late-type WC Wolf Rayet stars and IR sources at the Galactic center have
been used to probe  interstellar silicates in the diffuse interstellar medium
\nocite{aitken_etal80,roche_aitken85} (Aitken \etal\ 1980; Roche \& Aitken
1985). The Galactic center lines of sight have the advantage of copious
extinction ($A_V\sim 30$ magnitudes), but the disadvantage of being a crowded  
field where emission and absorption components can complicate the analysis of
interstellar features.  Late-type carbon-class (WC 8-9) Wolf-Rayet stars that
are extincted by 6 to 12 magnitudes of interstellar visual extinction, provide
an opportunity to study interstellar absorption features within the solar
circle.  Although these stars are known to have circumstellar dust, very
little of the visual extinction arises in the circumstellar environment
\nocite{roche_aitken84} (Roche \& Aitken 1984).

In this paper, we present the 2.38 to 40 \micron\ spectra from the Infrared
Space Observatory's (ISO) Short Wavelength Spectrometer (SWS) of four WC-type
Wolf-Rayet stars.  The data reduction is discussed in \S2.  In \S3 we describe
our method of fitting the spectral energy distribution of the WR stars. This
estimated continuum is used to extract the silicate absorption features.  In
\S4 we outline the calculations for the extinction profiles.  These calculated
profiles are compared to the best S/N observed profile in \S5.    Section 5
also compares our observed diffuse ISM profiles to observed silicate features
in different environments.     Section 6 gives an overview of the evolution of
silicates from their production sites to their incorporation into young star
environments.  In \S7 we discuss abundance constraints for the silicate
minerals considered and in \S8 we discuss other  possible contributors to the
absorption. Extinction curves from 1.25 to 25 \micron\ are constructed for the
``local'' interstellar medium and the Galactic center in \S9.  Finally, we
summarize our results in \S10.

\section{Observations and Data Reduction}
We present the 2.38-40  \micron\ spectra of the four late-type WC stars, WR
112, WR 118, WR 104 and WR 98a from the Short-Wavelength Spectrometer (SWS, de
Graauw \etal\ 1996) on the Infrared Space Observatory (ISO, Kessler \etal\
1996).  These spectra were acquired with the Astronomical Observing Template
(AOT) 01 speed 2 with an average resolving power ($\lambda/\Delta\lambda$)
between 250 and 600, and were first presented by \nocite{vanderhucht_etal96}
van der Hucht \etal\ (1996).  ISO-TDT numbers are given in Table~1.

Off-line processing (OLP) version 10.1 data were reduced starting at the
Standard Processed Data (SPD) level using IA$^3$.  For each band, the twelve
individual detector scans were normalized (``flat-fielded'') to a constant or a
first order polynomial.  Points that deviated more than 3 sigma from the
average were cut, and the resulting spectra were then rebinned to the average
resolving power for each band \nocite{leech_etal02} (Leech et al. 2002).  Band
2C (7--12\micron) spectra suffer from residuals caused by incomplete removal of
the relative spectral response function.  For these data, an extra reduction
step using the ``fringes'' routine in IA$^3$ was carried out after rebinning
the individual detector scans.  The WC spectra have low flux at long
wavelengths, so band 4 data (29--45 \micron) have low signal-to-noise and have
not been included in the final spectra shown in Fig.~\ref{fig:model}.

\section{Analysis}

The spectral energy distributions of the late-type WC stars are complicated  by
emission from heated carbon dust arising in their circumstellar environment
\nocite{vanderhucht_etal96,williams_vanderhucht_the87} (e.g., Williams, van der
Hucht \& Th\'e 1987; van der Hucht \etal\ 1996).  Modeling their spectral
energy distribution in detail is beyond the scope of this paper. Instead, we
will adopt  a physical, albeit simplistic, model for their continuum dust
emission in order to extract the interstellar silicate absorption features.

Following Williams \etal\ (1987), we assume that the WR stars' mass loss can
be modeled by optically thin spherical shells of amorphous carbon dust.   We
assume that the heating of the dust is dominated by visual/UV photons from the
central star and that the dust absorbs these with unit efficiency. The IR
emission efficiency, $Q\left(\nu\right)$, of carbon dust is well-represented by
a $\beta$-law with $\beta=1$  (e.g.,  $Q(\lambda) =
Q_o(\nu/\nu_o)^{\beta}$; Bussoletti \etal\ 1987, Mennella \etal\ 1995).  The
temperature is then described by,\nocite{bussoletti_etal87,mennella_etal95}
\begin{equation}
T_d(r) = T_o \left(\frac{r_o}{r}\right)^{2/(4+\beta)}
\end{equation}
where $T_o$ is the temperature at the inner radius $r_o$.    The contribution
to the flux by a shell is given by
\begin{equation}
F_{\nu}\, =\, 4\pi a^2\, \frac{4\pi r^2\, n\left(r\right)}{4\pi d^2}\, Q\left(\nu\right) \pi B_\nu\left(T_d\right)
\end{equation}
For the grain density distribution, we will adopt a power-law density distribution, $n\left(r\right) = n_o(r/r_o)^{-\alpha}$.
The flux is then given by,
\begin{equation}
F(\nu)\,=\, \frac{\left(4\, +\, \beta\right)\, n_o\, r_0 4\pi^2 a^2 Q_o \nu^{3+\beta-\gamma}}{d^2\, \nu_0^{\beta}\, c^2}\mathcal{F}\left(y_o, y_1\right)
\end{equation}
with $\gamma= \left(3-\alpha\right)\left(4+\beta\right)/2$ and $y= h\nu/kT_d$.
$\mathcal{F}$ is the integral,
\begin{equation}
\mathcal{F}\left(y_o, y_1\right)\, =\, \int_{y_o}^{y_1}\, \frac{y^{\gamma-1}}{\left(\exp\left[y\right]\, -\, 1\right)}\, dy
\end{equation}
where the integral limits apply to the inner and outer boundary of the shell.
For a constant outflow $\alpha=2$ and $\beta=1$, we have $\gamma=5/2$ When
$y_o$ is very small and $y_1$ is very large, the integral approaches the Rieman
zeta function and is independent of the exact boundary values. The frequency
dependence of the flux is then contained within the pre-integral factor,
$3+\beta-\gamma= 3/2$.

There are four parameters in the fit, $\alpha$, $\beta$, $T_o$, and $T_1$. We
have adopted $\alpha=2$ and $\beta=1$ and allowed $T_o$ and $T_1$ to vary, then
evaluated the integral. We have used this model to ``fit''  continua to the
observed spectra. The values of the parameter fits for each of the WR star
spectra are given in Table~2. The model ``fits'' are shown in
Fig.~\ref{fig:model} and the resulting optical depth spectra are shown in
Fig.~\ref{fig:taus}.  The optical depths, their ratios and the visual
extinctions (from van der Hucht 2001) are listed in Table~3. The lower panel in
Fig.~\ref{fig:taus} shows the spectra normalized to one at their peak in order
to compare the 9.7 \micron\ profiles.   The 9.7 \micron\ profiles are similar 
from 8 to 11.3 \micron, while there is some variation in the absorption between
features, from about 11.3 to 15.5 \micron, with the spectra of WR 104 and WR
118 being the outliers.   The central wavelength of the 18 \micron\ feature and
the ratio of the features is similar for 3 out of 4 sources.   The outlier, in
terms of the ratio of the silicate features, is the spectrum of WR 104.  This
could be due to uncertainties in the continuum that are a result of low S/N
data for $\lambda>25$ \micron, rather than a physical difference in the
environment along the line of sight.

\section{Calculating the Silicate Extinction Profiles \label{calculations}}

In order to compare the laboratory derived silicate minerals to the observed
absorption spectrum, one needs to calculate extinction profiles from the
available optical constants.  We assume that the size of the silicate grains is
much smaller than the wavelength (Rayleigh limit). For spherical particles in
the Rayleigh limit, the absorption efficiency can be approximated as
\begin{equation}
Q_{abs} = 4x \Im \left\{\frac{\epsilon -1}{\epsilon + 2}\right\}
\end{equation}
where $x=2\pi a/\lambda$ and $a$ is the particle size.  The cross section per
unit volume  for absorption is
\begin{equation}\label{eq:sphere}
\frac{C_{abs}}{V} = \pi a^2 Q_{abs} =
\frac{6\pi}{\lambda}\Im\left\{\frac{\epsilon-1}{\epsilon+2}\right\}
\end{equation}
For a continuous distribution of ellipsoids (CDE), assuming that all shapes are
equally probable, the average absorption cross section per unit volume is
\begin{equation}\label{eq:cde}
\frac{<C_{abs}>}{V} = \frac{2\pi}{\lambda}\Im\left\{ \frac{2\epsilon}{\epsilon
-1}\ln\epsilon\right\}
\end{equation}
The above equations are valid for uniform solid particles.
To calculate the dielectric function for silicates with inclusions, we used  the
Bruggeman rule \nocite{bohren_huffman83} (Bohren \& Huffman 1983):
\begin{equation}
f\frac{\epsilon-\epsilon_{\rm ave}}{\epsilon+2\epsilon_{\rm ave}}
 + (1-f)\frac{\epsilon_m-\epsilon_{\rm ave}}{\epsilon_m+2\epsilon_{\rm ave}}
 = 0
\end{equation}
where $\epsilon_{\rm ave}$ is the average dielectric constant to be substituted
into equations~\ref{eq:sphere} or \ref{eq:cde}, $\epsilon_m$ is the dielectric
function of the matrix (silicate) and $f$ is the filling factor.  For grains
will high porosity, the filling  factor ($f$) above is replaced with the
correction derived by Ossenkopf (1991)  for spherical  inclusions
\nocite{ossenkopf91}:
\begin{equation}
f_{\rm spheres} = f\cos^2(\pi f).
\end{equation}

The cross-sections per unit volume ($C_{\rm abs}/V$) for the silicate minerals
and shapes considered here are listed in Table~4.

\section{The Silicate Features}

\subsection{The Profile of the Interstellar Silicate Features}

In this section we will compare the 9.7 and 18 \micron\ profiles observed
toward the WC stars, representative of  (local) diffuse ISM dust, with
silicate profiles in different environments.

\subsubsection{The $\mu$ Cephei profile}
The emissivity curve for the red supergiant $\mu$ Cephei has historically been
used to represent and model interstellar extinction in the 8--13 \micron\
region \nocite{roche_aitken84} (e.g., Roche \& Aitken 1984).  The silicates
around $\mu$ Cep were presumably formed in the outflow of the star.   We
have re-examined the $\mu$ Cep spectrum in light of the identification of
photospheric lines in the full ISO-SWS spectrum \nocite{tsuji00} (Tsuji 2000). 
The photospheric model by Tsuji (2000) shows that an absorptive SiO component,
whose onset is just shortward of the 9.7 \micron\ silicate emission feature,
partially underlies it.   We first subtract the stellar continuum represented
by a 3600 K blackbody curve (Fig.~\ref{fig:mucep}, left panel).  The resulting
excess emission is fitted with a 250 K blackbody curve, representing the
underlying dust emission.  An emissivity curve is created by dividing the $\mu$
Cep emission spectrum by the 250\,K blackbody curve.  The final result is shown
in the right panel of Fig.~\ref{fig:mucep} along with the absorption spectrum
of WR 98a.  The $\mu$ Cep emissivity curve provides a reasonable representation
of the diffuse ISM absorption feature. 

\subsubsection{The Galactic Center Quintuplet Profile}
The line of sight toward the Galactic center has long been used to probe the
properties of diffuse ISM dust.  Studies of the ISM silicate absorption
features toward the central sources are complicated by dust emission
\nocite{smith_aitken_roche90,roche_aitken85} (Roche \& Aitken 1985; Smith,
Aitken \& Roche 1990). Northeast of the proper center is the Quintuplet
Cluster, whose bright sources have also been used to study general ISM dust
characteristics.  The nature of the IR sources remains enigmatic, however it is
thought that they could be dusty WC stars
\nocite{figer_mclean_morris99,chiar_tielens01,moneti_etal01} (Figer \etal\
1999; Chiar \& Tielens 2001; Monieti \etal\ 2001), much like the sources
discussed in this paper. Spectropolarimetry of a few of the brightest IR
sources (GCS 3-II, III, IV) in the Quintuplet Cluster shows that these lines of
sight are dominated by pure absorptive polarization across the 9.7 \micron\
silicate  feature \nocite{smith_etal00} (Smith \etal\ 2000), making them
excellent probes of the ISM absorption feature. 

Here we use the Infrared Space Observatory's Short Wavelength Spectrometer
spectrum of GCS 3 to compare the silicate absorption profiles toward the
Quintuplet with those seen toward WR 98a.  We derived the optical depth
spectrum for GCS 3 by fitting a fourth order polynomial to the flux spectrum
(Fig.~\ref{fig:ohir_gcs3}, top). Fig.~\ref{fig:ohir_gcs3} (bottom) shows that
the relative depths of the 9.7 and 18 \micron\ silicate features are similar
for GCS 3 and WR 98a.  The spectra deviate from each other most noticeably in
the wavelength region between the two silicate absorption features.  This could
be a result of the difficulty in defining the continuum for the WR 98a spectrum
longward of the 18 \micron\ feature.  Otherwise, the spectra have similar
characteristics, including the $\tau_{18}/\tau_{9.7}$ ratio, implying a similar
silicate mineralogy for the local ISM (WR 98a) and the Galactic center (GCS 3).

\subsubsection{The Asymptotic Giant Branch Star Profile}

Interstellar silicates are thought to be formed through condensation in the
outflows of Asymptotic Giant Branch (AGB) stars.  As an example of the profiles
of silicates formed in such environments, we compare the 9.7 and 18 \micron\
profiles of silicates obtained for diffuse ISM with those derived from
observations of OH-IR 127.8+0.0 (Fig.~\ref{fig:ohir_gcs3}, bottom). The  {\it
model\/} 9.7 \micron\ extinction profile for the oxygen-rich AGB star OH-IR
127.8+0.0 (Kemper \etal\ 2002) is broader on the red wing compared to the ISM
(WR 98a and GCS3 I) profiles.  This additional absorption has been attributed
to the presence of crystalline silicates \nocite{demyk_etal00} (Demyk \etal\
2000).   The 18 \micron\ OH-IR star silicate absorption feature peaks at 17.5
\micron\ (Demyk \etal\ 2000; Kemper \etal\ 2002), shortward of the diffuse ISM
feature which peaks at 18.5 \micron.  Demyk \etal\ (2000) suggest that the 
difference in peak wavelength of the 18 \micron\ feature is due to the
relatively low pyroxene-glass content of the OH-IR star silicates compared to those
in the ISM (Demyk \etal\ 2000).

\subsubsection{The Orion profile}
The inner regions of the Orion HII region (Trapezium region) show a 9.7
\micron\ silicate emission feature \nocite{forrest_gillett_stein75} (Forrest,
Gillett, \& Stein 1975) that is similar to the absorption feature observed in
the cold dense interstellar medium \nocite{gillett_etal75} (e.g., molecular
cloud sources; Gillett \etal\ 1975). This 9.7 \micron\ profile has been used
by  Draine \& Lee (1984; DL) as constraints on their model of interstellar
silicates.  Figure~\ref{fig:draine} shows that the DL ``astronomical'' silicate
profile does not match the observed central wavelength or width of the 9.7
\micron\ absorption. The observed diffuse ISM 9.7 \micron\ profile peaks
longward of the Trapezium profile. Draine and Lee (1984) did not attempt a
detailed fit of an observed ``20'' $\mu$m silicate feature but only constrained
it to peak at 18$\mu$m and to be $\simeq$40\%\ of the 9.7 $\mu$m feature in
strength. The feature in the diffuse interstellar medium peaks at slightly
longer wavelengths and is slightly stronger (deeper) than the DL feature.

\subsubsection{The Dense Interstellar Medium Profile}

The 9.7 \micron\ Trapezium profile has been shown to be more similar to the
absorption feature observed in dense clouds rather than that in the diffuse
ISM. \nocite{bowey_adamson_whittet98,whittet_etal88} (e.g., Whittet \etal\
1988; Bowey, Adamson \& Whittet \etal\ 1998).  Bowey \etal\ (1998) ascribe the
differences in the diffuse ISM and dense cloud silicate absorption profiles to
excess absorption, due to a non-volatile grain component,  on the
long-wavelength wing of the dense cloud feature.  Similarly,
\nocite{demyk_etal99} Demyk \etal\ (1999) find that for deeply embedded massive
protostars, which probably also have dense cloud dust along the line of sight,
the silicate absorption profile peaks shortward of that in the diffuse ISM and
has additional absorption on the long wavelength wing.  However, for these
sources, absorption in the long-wavelength wing is attributed to simple ice
species \nocite{demyk_etal99} (Demyk \etal\ 1999).  This explanation for the
profile difference between the diffuse ISM as measured toward the WC stars and
the Galactic center versus the dense cloud profile would not work well for the
Trapezium region.  First, volatile ices are not expected to survive long in the
ionized gas of the HII region.  In addition, this dust has to be fairly warm in
order to emit at 10 \micron\ and ice already evaporates at $\sim 100\,$K in the
ISM.

\subsection{Comparison with Laboratory Derived Profiles}
The key to identifying the silicate mineral(s) responsible for the observed
absorption features is finding the combination of mineral(s), shape and
porosity that reproduces the observed peak wavelengths, widths and relative
depths of the observed features.  Laboratory studies show that the profiles of
the silicate features depend on the composition and structure of the silicate
material.  We will focus here on the results for amorphous silicates since it
has been shown that crystalline silicates are not prevalent in the diffuse ISM
(Kemper, Vriend \& Tielens 2004). Silicates are an important component of
interstellar dust accounting for a major fraction of the interstellar dust
volume. Given the observed interstellar silicate dust volume, we will focus on
silicates based upon the abundant elements, magnesium and iron (see also \S7).
We will only consider  here magnesium-iron silicates with an olivine or
pyroxene stoichiometry. The general chemical formulae for olivines and
pyroxenes are  [(Mg$_2y$,Fe$_{2-2y}$)SiO$_4$] (with $y$ between 0 and 1) and
[Mg$_x$,Fe$_{1-x}$SiO$_3$], respectively. Generally, it is thought that
interstellar silicates are formed in the ejecta of Asymptotic Giant Branch
stars through reactions with gaseous species \nocite{salpeter77} (Salpeter
1977).  Magnesium silicate compounds are calculated to be the first major
condensates in a gas with solar system composition cooling down from a high
temperature (Grossman \& Larimer 1974). For the pressures relevant to these
outflows, at slightly lower temperatures, iron is then expected to react with
these magnesium silicates rather than condense as separate iron grains, forming
iron-magnesium silicates.  While the magnesium silicates are formed at
relatively high temperatures and may well have a crystalline structure - after
all, crystalline forsterite (Mg$_2$SiO$_4$) and enstatite (MgSiO$_3$) are
observed to be abundant in these stellar ejecta, the reaction with gaseous iron
occurs below the glass temperature of these compounds  and, hence, due to
kinetics, an amorphous structure will result 
\nocite{tielens_allamandola87,tielens90} (Tielens \& Allamandola 1987; Tielens
1990). While strictly speaking the terms olivines and pyroxenes are reserved
for crystalline silicates with definite mineral structures and hence not
appropriate for amorphous silicates, for ease of discussion, we will loosely
call these amorphous compounds ``olivine'' and ``pyroxene'' glasses in the
subsequent sections.   Because of their formation history, these silicates
likely have the stoichiometry corresponding to olivines and pyroxenes not only
on a global level but also on a microscopic level and the amorphous nature of
these grains reflects small variations in the  nearest-neighbor bond angle
and/or bond distance which destroys all long range order.  

The 10 \micron\ feature in olivine-glass silicates peaks at somewhat longer
wavelengths than that for pyroxene glasses.  At 18 \micron, the olivine-glass
feature is somewhat sharper and peaks at somewhat shorter wavelengths than the
pyroxenes.  In addition, for both minerals, the extinction profile of the 9.7 and
18 \micron\ features broadens and  the strength of the 18 \micron\ feature
relative to the 9.7 \micron\ feature increases as the  Mg:Fe ratio decreases
\nocite{dorschner_etal95} (Dorschner \etal\ 1995).  The Fe:Mg ratio ranges from 0
to 1.5 and 1 to 1.5 for the pyroxene and olivine glasses, respectively,
considered by Dorschner \etal\ (1995). The extinction profiles also depend on the
shape of the absorbing grains and the porosity of the grains.  In the remainder
of this section, we will make a detailed comparison between the observed silicate
profiles and theoretically calculated profiles based upon the laboratory measured
optical properties \nocite{henning_etal99} (Henning \etal\
1999).\footnote{Optical constants for amorphous olivine and pyroxene glasses from
the Laboratory Astrophysics Group of the Astrophysical Institute and University
Observatory in Jena
(http://www.astro.uni-jena.de/Laboratory/Database/odata.html)}

\subsubsection{Solid Spheres}

Amorphous olivine-glass is generally used to represent the ISM silicate absorption
feature at 9.7 \micron.  However, while amorphous olivine-glass spheres provide a
reasonable match to the 9.7 \micron\ absorption feature, the observed 18.5
\micron\ feature is not well matched in either central wavelength or relative
depth.  As shown in Fig.~\ref{fig:spheres} (top left panel), the spectrum of
olivine-glass spheres peaks near 17 \micron\ and is too deep relative to the observed
feature.  The spectra of amorphous pyroxene-glass spheres peak near 9.3 \micron\ and
therefore do not provide a good match to the observed 9.7  \micron\ feature.  On
the other hand, the relative depths of the 9.7 and 18 \micron\ features, as
well as the central wavelength of the 18 \micron\ feature are well-reproduced by
amorphous pyroxene glasses (Fig.~\ref{fig:spheres}, top left panel).  

Thermodynamic equilibrium calculations show that, for temperatures less than
about 1300 K\footnote{The exact temperature where this conversion sets in depends
somewhat on the pressure.}, solid forsterite is converted into solid enstatite
with excess gaseous SiO \nocite{grossman_larimer74} (Grossman and Larimer 1974).
The extent to which this reaction proceeds before grain formation freezes out in
the stellar ejecta depends on the  timescale and density/temperature structure in
the outflow and, hence, is difficult to constrain observationally. For this
reason, we attempted to reproduce the astronomical features by a mix of these two
components (assuming separate populations).  The bottom left panel in
Fig.~\ref{fig:spheres}  shows that the match using an arbitrary
pyroxene-glass/olivine-glass mixture is improved over  considering only one
silicate species.  However, the match between the pyroxene-glass/olivine-glass
``mixture'' and the observed spectrum is less than ideal between 11 and 15
\micron.  In addition,  the peak wavelengths of both features are somewhat
blueward of the observed feature.  One way to shift the peak wavelengths redward
is to incorporate vacuum into the grains.  We discuss such porous grains below.

\subsubsection{Porous Spheres}
Observations of dust in dense clouds suggest that grain growth is dominated by
coagulation \nocite{jura80} (Jura 1980), thus grains are likely to be porous. 
Porous grains have larger extinction cross sections relative to their solid
counterparts and therefore require less of a given element to account for,
e.g., the silicate interstellar absorption features.  Mathis (1998)
\nocite{mathis98} argues that the cosmic abundance requirements can be met if
25\% of the grain volume is vacuum. An upper limit of 60\%\ on the porosity is
given by the width of the interstellar polarization law
\nocite{wolff_clayton_meade93} (Wolff, Clayton, \& Meade 1993).  

We used the Bruggeman rule (Bohren \& Huffman 1983) to calculate the extinction
profiles for porous silicates, assuming that the vacuum is in the form of
spherical inclusions (\S\ref{calculations}). In general, increasing the vacuum
volume has the effect of decreasing the 9.7 \micron\ to 18 \micron\ ratio and
shifting the peak absorption of both bands redward. Fig.~\ref{fig:spheres} (top
right panel) demonstrates how the extinction profiles of pyroxene-glass and
olivine-glass spheres are modified by 50\% vacuum.  pyroxene-glass spheres that are
50\% porous do a fair job at reproducing the astronomical spectrum, although they
provide too much extinction on the blue side of the 9.7 \micron\ feature.  Porous
olivine-glass spheres (50\% vacuum) do not reproduce the observed spectrum, the
biggest problem is the 18 \micron\ to 9.7 \micron\ band ratio that is too high to
match the observations.  The extinction profile of a mixture of porous (50\%
vacuum) olivine-glass and pyroxene-glass spheres is shown in the bottom right panel of
Fig.~\ref{fig:spheres}.  In this case, the 9.7 \micron\ band is somewhat broader
than the observed feature and the 18.5 \micron\ to 9.7 \micron\ band ratio is
overestimated, although not significantly.

\subsubsection{Continuous Distribution of Solid Ellipsoids}
A continuous distribution of ellipsoids (CDE) where each shape occurs with 
equal probability could reflect the characteristic grain population in the
ISM.   Figure~\ref{fig:cde} (left panels) shows the calculated extinction
profiles, using equation \ref{eq:cde},  for solid CDE. The
extinction profiles for CDE are somewhat broader and are shifted redward
relative to those for spheres.    As shown in Fig.~\ref{fig:cde}, the
extinction profile for olivine-glass CDE is broader on the red side of the 9.7
\micron\ profile, and produces a stronger 18.5 \micron\ feature, relative to
the observed profile.  On the other hand, the profiles calculated for pyroxene-glass
CDE match the relative strengths of the 9.7 and 18.5 \micron\ features well,
but are broader than the observed 9.7 \micron\ profile on the blue side. 
Co-adding the extinction profiles of  pyroxene-glass and olivine-glass CDE improves the
match to the wings of the 9.7 \micron\ profile, however, the relative strength
of the 9.7 \micron\ feature is underestimated (Fig.~\ref{fig:cde}, bottom left
panel).

\subsubsection{Continuous Distribution of Porous Ellipsoids}
We use the Bruggeman rule again to calculate the extinction profile for porous
CDE with 50\% porosity.  The extinction profiles for porous CDE result in a
somewhat increased 18 \micron\  to 9.7 \micron\ feature ratio compared to solid
CDE.  In addition, both profiles are broader than those for solid CDE.

\subsubsection{Aluminosilicates}

Aluminum is heavily depleted in the interstellar medium \nocite{whittet_bk03}
(Whittet 2003), and as such, is likely to be incorporated into silicate dust. 
Aluminosilicates were considered for circumstellar silicates around oxygen-rich
stars because they provide much-needed opacity in the trough region between the
9.7 and 18 \micron\ features (Mutschke \etal\ 1998).  We considered Fe- and
non-Fe- containing aluminosilicates for the diffuse ISM absorption features. 
Extinction profiles for porous and solid spheres and CDE were compared with the
observations based on the optical constants by Mutschke \etal\ (1998).  Fitting
the observed silicate profiles with aluminosilicates (with and without Fe) with
Si:Al equal to 2:1 requires more Al than available in the ISM independent of
grain shape and porosity.  The Fe-containing aluminosilicates provide more
opacity in the trough region compared to aluminosilicates with no Fe.  In
general, for  the  alumino-silicates considered, the relative depths of the
observed absorption features are matched by the extinction profiles for porous
spheres, or solid or porous CDE; solid spheres result in a larger 9.7\micron/18
\micron\ ratio than observed.  The observed 18 \micron\ profile shape is also
well-matched by the extinction profiles for porous spheres, or solid or porous
CDE.  Aluminosilicates look less promising in the spectral region of the
observed 9.7 \micron\ absorption feature.  Here, the calculated extinction
profiles do not provide a good match to the observed peak wavelength and the
profile width, simultaneously.  Thus, it is unlikely that aluminosilicates can
account for the observed silicate profiles on their own, but they may still be
present as a component of the interstellar silicate dust.

\subsubsection{Iron Oxides}
Theoretical calculations under conditions appropriate for the cooling solar
nebula (e.g., high pressures as compared to stellar ejecta) predict that iron
oxide forms by oxidation of metallic iron with excess H$_2$O at about 600 K
(Barshay \& Lewis 1976). The fate of condensing iron is less clear in the ejecta
of asymptotic red giant branch stars. Depending on temperature, pressure, and
oxygen partial pressure, iron condensation can shift from metallic iron, to
iron-magnesium silicates, to ferrous iron (FeO, wuestite), mixed ferrous-ferric
oxide (Fe$_3$O$_4$; magnetite), to ferric iron (Fe$_2$O$_3$; hematite) in
thermodynamic equilibrium calculations. Of course, kinetic factors may hamper the
formation of some of these compounds (eg., metallic iron). In interstellar and
circumstellar medium settings, the simplest iron oxide, ferrous oxide (FeO), has
at times been considered as an important contributor to the extinction in the
infrared \nocite{henning_etal95,demyk_etal00,kemper_etal02} (Hennning \etal\
1995; Demyk \etal\ 2000; Kemper \etal\ 2002). The presence of circumstellar
magnesio-wuestite grains has been inferred from the 19 \micron\ emission feature
in the spectra of AGB stars with low mass loss rates
\nocite{cami02,posch_etal02,heras_hony05} (Cami 2002; Posch \etal\ 2002; Heras \&
Hony 2005). However, those grains are considered to be largely magnesium-rich. 
On Earth, wuestite (FeO) is considered an important constituent of the lower
mantle and natural samples of wuestite are available. In the laboratory, wuestite
is routinely made using a gel method and its optical properties have been
measured for astronomical purposes. Of course, in the presence of excess oxygen,
ferrous oxide burns explosively at temperatures exceeding ~600 K, but - in view
of kinetic considerations - that may not be very relevant for interstellar
conditions (where temperatures seldomly reach such high values), anyway. Here, in
view of these discussions in the literature, we merely examine the effect of
small FeO inclusions in silicate grains on the 9.7 and 18 \micron\ silicate
profiles if they were present in the ISM. 

Spherical inclusions of FeO are considered here for interstellar silicates
using the Bruggeman rule.  As discussed by \nocite{henning_etal95} Henning
\etal\ (1995), the presence of FeO in silicates has the affect of increasing
the 18\micron/9.7\micron\ ratio and broadening both bands.   The 9.7 \micron\
and 18 \micron\ features shift blueward and redward, respectively, as the
volume of FeO inclusions is increased.  Thus, adding FeO (spherical) inclusions
to solid olivine-glass or pyroxene-glass spheres  would worsen the match to the
interstellar spectrum. For solid olivine-glass spheres, FeO inclusions increases the
depth of the 18 \micron\ band, and for solid pyroxene-glass spheres, the 10 \micron\
band is shifted to even shorter wavelengths.  Similarly, adding FeO to porous
spheres does not improve the fit. Therefore, iron oxides are probably not a
significant grain component in ISM dust.

\subsubsection{Summary}

Considering the parameter space investigated here, we find two equally good
matches to  the 9.7 and 18 \micron\ silicate absorption profiles. Our criteria
for a good match are as follows: the extinction profile should fit  the relative
depths, the central wavelengths, and the wings of both the 9.7 and 18 \micron\
absorption features.  One good match is achieved by co-adding the extinction
profiles of olivine-glass and pyroxene-glass spheres with 50\% porosity.  An equally good
match is given by the co-addition of pyroxene-glass and olivine-glass solid CDE.  Both these
fits require a greater contribution by mass from pyroxene-glass relative to olivine-glass.
This differs from the conclusion reached by Kemper, Vriend \& Tielens (2004) that
olivine glasses account for the majority of the mass of diffuse ISM silicates. 
However, these authors consider only the 9.7 \micron\ feature which we also find
is best matched with olivine glasses if considered by itself.

Several of the calculated extinction profiles shown result in unsatisfactory
fits to the observed silicate profiles.   While the extinction profile for
solid olivine-glass spheres matches the observed 9.7 \micron\ silicate profile well,
it does not match the central wavelength or relative depth of the 18 \micron\
feature (Fig.~\ref{fig:spheres}, top left).  On the other hand, solid pyroxene-glass
spheres match the 18 \micron\ feature, but not the 9.7 \micron\ feature. 
Co-adding the pyroxene-glass and olivine-glass profiles for solid spheres does not improve
the fit (Fig.~\ref{fig:spheres}, bottom left). Fig.~\ref{fig:spheres} (top
right) shows that adding porosity improves the fit for pyroxene-glass spheres, though
the blue wing of the 9.7 \micron\ feature is not well-matched.  Porous olivine-glass
spheres do not match either the 9.7 or 18 \micron\ profiles.    Neither the solid CDE  or porous CDE
extinction profiles for pyroxene or olivine glasses on their own result in
satisfactory fits to the observed silicate features (Fig.~\ref{fig:cde}).

Aluminosilicates are good candidates for the opacity in the trough region
between the two silicate absorption features.  Many of the  calculated
extinction profiles also match the 18 \micron\ profile  and the relative depths
of the features well.  However, these minerals do not provide a good match to
the 9.7 \micron\ profile shape or peak wavelength.  Thus, we conclude that
these minerals are unlikely to account for the observed features on their own,
but could partially contribute to the observed absorption.  We also find that
the presence of iron oxides cannot be accounted for in the observed ISM
spectrum.

\section{The Lifecycle of Interstellar Silicates}
The lifecycle of interstellar silicates starts with the nucleation and chemical
growth at high densities and temperatures in the ejecta from stars such as
Asymptotic Giant Branch stars and supergiants. This ejected material is rapidly
mixed with other gas and dust in the interstellar medium. In the interstellar
medium, dust cycles many times between the intercloud and cloud phases until it
either is destroyed by fast ($\sim100$ km/s) supernova shocks or is incorporated
into newly formed stars or planetary systems.

Infrared observations with the Infrared Space Observatory have revealed that
crystalline silicates are abundant ($\sim15$\% by volume) in the initial stages
-- the ejecta from stars -- as well as in the last stages of its life cycle --
circumstellar disks around Herbig AeBe stars and T-Tauri stars --  but are
absent in the middle portion of this evolutionary scenario
\nocite{sylvester_etal99,kemper_vriend_tielens04,malfait_etal99,waters_waelkens98} 
(Sylvester \etal\ 1999; Kemper \etal\ 2004, 2005; Malfait \etal\ 1999; Waters
\& Waelkens 1998). Indeed, the smooth profile of the 9.7 \micron\ silicate
feature implies that the crystalline-to-amorphous silicate fraction is less
than 1\% in the interstellar medium (Kemper \etal\ 2004, 2005). Likely, the
crystalline silicates injected in to the interstellar medium by stars are 
amorphized by high energy galactic cosmic rays on a timescale
\nocite{bringa_etal05} (70 million years; Bringa \etal\ 2005) which is short
compared to interstellar dust evolution timescales \nocite{jones_etal96}
($\sim500-4000$ million years; Jones \etal\ 1996). The observed high abundance
of crystalline silicates in  circumstellar disks surrounding young stars must
then reflect subsequent processing of grains in these environments. Recent
mid-infrared spectroscopic interferometry on AU-scale-sizes has revealed the
presence of a strong gradient in the crystallinity of silicates in the disks of
the three objects investigated \nocite{vanboekel_etal04} (van Boekel \etal\
2004). Likely this reflects the rapid annealing in the hot inner regions of
these disks coupled with turbulent diffusion outward of the crystalline
grains. 

\section{Elemental Abundances}

The fraction of the elements locked up in interstellar dust grains provides a
key test for any realistic dust model \nocite{snow_witt96,mathis96} (e.g. Snow
\& Witt 1996; Mathis 1996).  Interstellar abundances have recently been
reviewed by \nocite{sofia_meyer01} Sofia \& Meyer (2001) who show that young
($\le2$ Gyr) F and G stars and the Sun provide the best representations of the
interstellar abundances. 

We calculate the silicate abundance per unit hydrogen atom for both solid  and
50\% porosity silicate (olivine-glass and pyroxene-glass) spheres and CDE, and tabulate the
results in Table~5. The final row in Table~5 lists the silicate abundance using
a mixed population of olivine-glass and pyroxene-glass grains. We assume an average total
hydrogen column density per unit extinction ($<N_{\rm H}/A_V>$) of
$1.9\times10^{21}$ cm$^{-2}$ mag$^{-1}$, with a typical scatter of less than
30\% over a wide range of $E(B-V)$ (Bohlin et al.~1978). We take
$A_V/\tau_{9.7}=18\pm1$, determined for the interstellar medium (Roche \&
Aitken 1984\nocite{roche_aitken84}).  Then, the abundance of silicate atoms
relative to hydrogen locked up in solid form is, 

\begin{equation}
\frac{N_{\rm silicate}}{N_{\rm H}} = \frac{1}{C_{abs}/V} \times
\frac{\rho_{\rm silicate}}{1.66\times10^{-24}M_{\rm silicate}} \times
\frac{1}{18.5\times 1.9\times10^{21}} 
\end{equation}

where $\rho_{\rm silicate}$ and $M_{\rm silicate}$ are the specific density 
and the atomic mass per silicon atom of the silicate material. Our discussion
on elemental abundances will be based on the evaluation by Sofia \& Meyer
(2001).  They conclude that the atomic Si available to be incorporated into
dust is between $3.44-3.99\times10^{-5}$ per H atom, enough to alleviate the Si
underabundance problem proposed by Snow \& Witt (1996), for many (but not all)
silicate dust models.  Using solid olivine-glass and/or pyroxene-glass spheres to account
for the observed silicate profiles requires 5 to 49\% more than the available
Si.  Similarly, modeling the silicate absorption features with solid pyroxene-glass
CDE requires somewhat more than the available Si.   Assuming that interstellar
silicate grains are porous allows the interstellar features to be accounted for
without using up the available Si.  There is enough available Mg and O to be
incorporated into silicate grains.  The elemental abundance of Fe, on the other
hand, poses a problem for olivine-glass grains unless porous grains are used
(Table~5).

Comparison of elemental abundances in the cold neutral medium as compared to
the warm neutral medium suggests that a fraction of 0.3 of the elemental
Si is locked up as a dust component which is not as refractory as silicate
materials and is more readily destroyed by interstellar shocks
\nocite{sembach_savage96,tielens97} (Sembach \& Savage 1996; Tielens 1997).  If
we assume that we only have $2.4 - 2.8\times10^{-5}$ Si per H-atom available
for interstellar silicate dust, all models would fail the test (Table 5). 
However, porous spheres and CDE models would come the closest.  The small
discrepancy may just reflect a small difference in the elemental abundance of
silicon in the direction of the inner Galaxy due to the elemental abundance
gradient of the galaxy.  From other elements, we estimate that this may amount
to an increase in the Si abundance of a factor of 1.2 over the typical
distance ($\sim$2 kpc) for which the local value of $A_V/\tau_{9.7}$ was 
determined.

\section{Other dust components}


Silicon carbide is produced in the outflows of carbon-rich low-mass  asymptotic
giant branch stars \nocite{anders_zinner93} 
\nocite{blanco_etal94,speck_barlow_skinner97,groenewegen95} (Blanco \etal\
1994; Groenewegen 1995; Speck \etal\ 1997) and is incorporated into meteoritic
dust (Anders \& Zinner 1993 and references therein).  It should be prevalent in
ISM dust, but limits of $<5$\% have been placed on its abundance
\nocite{whittet_duley_martin90} (Whittet, Duley, \& Martin 1990). 

There is no evidence for SiC grains in the observed spectra. We put a limit on
the amount of SiC present in the ISM using the (uncorrected) extinction
efficiencies given by Borghesi \etal\ (1987) for $\beta$-SiC since it peaks
near the observed carbon-star emission feature.  Its extinction efficiency is
($Q_{\rm ext}/a)_{\rm max} = 3.32\times10^4$ cm$^{-1}$.  Assuming a specific
density $s=3.2$ g cm$^{-3}$, the mass absorption coefficient, $\kappa$ is
calculated to be 7780 cm$^2$ g$^{-1}$.  To calculate the column density of Si
atoms contained in SiC grains, we use the equation (Whittet  \etal\ 1990)

\begin{equation}
N_x({\rm Si}) = \frac{f_\lambda \tau_\lambda}{28m_{\rm H}\kappa_\lambda}
\end{equation}

where $f_\lambda$ is equal to 0.7, the mass-fraction of Si in SiC. The optical
depth is calculated from the spectra of WR 98a and WR 112.  We do not use the
WR 104 spectrum since, at wavelengths longer than 11 \micron, it significantly
deviates from the mean of the other 3 spectra.  For WR 118,  olivine-glass and
pyroxene-glass already  slightly overcompensate for the absorption around 11.3
\micron.  

We use a $\chi^2$-fitting routine to determine the maximum allowable depth of
$\beta$-SiC in the WR 98a and WR 112 spectra using a three-component fit
($\beta$-SiC, olivine-glass and pyroxene-glass solid spheres).  We determine $\tau_{11.3}
<$ 0.016 and 0.019 for WR 98a and WR 112, respectively. Thus, the limit on the
fraction of Si in SiC with that in silicates is 0.03 and 0.04 for WR 98a and WR
112, respectively.  The results of our calculations are tabulated in Table~6.

\section{Extinction Curve}

The extinction curve in the infrared is generally accepted to be uniform
between 0.9 and 5 \micron.  Based on an extensive compilation of photometric
data for dense clouds and diffuse ISM sightlines, Martin \& Whittet  (1990)
suggest a universal interstellar extinction curve from 0.35 to 5 \micron,
represented by a power law, $A_{\lambda}/A_V = \lambda^{-1.8}$.  Recently,
\nocite{indebetouw_etal05} Indebetouw \etal\ (2005) used photometric data from
1.25 to 8 \micron\ of the Galactic plane from both the Spitzer-GLIMPSE program
and 2MASS to investigate the interstellar extinction curve and extend its
validity to 8 \micron.  They deduce the extinction law, $\log[A_\lambda/A_K] =
0.61-2.22\log(\lambda) + 1.21[\log(\lambda)]^2$, that is valid for the dense
and diffuse ISM within the solar circle.  These authors show that their
extinction curve provides a reasonable match to the Lutz (1999) Galactic center
(GC) values that were deduced from hydrogen recombination lines, demonstrating
that, at least in the near-IR, the extinction in the local ISM and the GC are
similar. We plot both data sets in Fig.~\ref{fig:extinction}, renormalizing the
Lutz (1999) points to the K-band extinction,  assuming $A_V=29$ magnitudes and
$A_K=3.28$ magnitudes (Figer, McLean \& Morris 1999). Given the good agreement
of the Indebetouw \etal\ (2005) and the Lutz (1999) extinction points, we
combined them and carried out a least-squares fit assuming the same functional
form as Indebetouw \etal\ (2005) and find $\log[A_\lambda/A_K] =
0.65-2.40\log(\lambda) + 1.34[\log(\lambda)]^2$.  We use this equation to
represent the continuum extinction, assuming that at $\lambda>8$ \micron, the
continuum remains constant and the silicate extinction dominates.

It is well-known that there is more silicate per unit of visual extinction
toward the GC compared to the local ISM \nocite{roche_aitken84,roche_aitken85}
(e.g., Roche \& Aitken 1984, 1985).   Thus, longward of 8 \micron, we need to
construct individual extinction curves for the local ISM and the GC.  To create
the ``GC extinction curve'' we first normalize the silicate optical depth
spectrum to the average visual extinction to silicate optical depth ratio,
$A_V/\tau_{9.7}=9$, determined by Roche \& Aitken (1985) for the Galactic
Center.  The spectrum was then normalized to $A_K$ using the Figer \etal\
(1999) $A_V$ and $A_K$ values for the Galactic Center (see above).  Finally,
this curve was added to our continuum extinction curve assuming that the
``continuum'' remains constant longward of 8 \micron.  The final GC extinction
curve is shown in Fig.~\ref{fig:extinction} (left panel) and partially tabulated
in Table~7.  A complete  downloadable version of the table is available in the
electronic edition of the Journal.

To create a complete extinction curve for the local ISM, we  carry out the same
set of steps as for the GC, but this time we use  $A_V/\tau_{9.7}=18$,
appropriate for the local ISM     (Roche \& Aitken 1984; Rieke \& Lebofsky
1985).  In order to normalize this to the extinction in the K band, we use
$A_K/A_V$ =  0.09 (Whittet 2004). Finally, we add the normalized WR 98a
silicate profile to the continuum extinction.   The complete local interstellar
extinction curve is  shown in Fig.~{\ref{fig:extinction} (right panel).  A
partial list of tabulated values is shown in Table~7; a complete downloadable
version of the table is available in the electronic edition of the Journal.

\section{Summary and Discussion}

In this paper, we examined the silicate absorption features observed in the
diffuse ISM toward WC-type WR stars.   As earlier work has shown, the observed
9.7 \micron\ feature is most similar to the $\mu$ Cephei profile, and differs
from the Trapezium profile.  The latter is used by Draine \& Lee (1984) to
compute their ``astronomical'' silicate profile, and is found to be more
similar to the silicate absorption features observed in dense clouds. We have
produced infrared extinction curves appropriate for the GC and local ISM.  These
two curves, which range from 1.25 to about 30 \micron\ are available for
download in the electronic version of the Journal. 

We find that pyroxene-glass silicates dominate by mass in  the diffuse ISM
silicates.  Pyroxene-glass silicates are also found to dominate the spectra of
YSOs \nocite{demyk_etal99,dorschner_etal95,jaeger_etal94} (Demyk \etal\ 1999;
Dorschner \etal\ 1995 and references therein; Jaeger \etal\ 1994). In contrast,
for evolved stars, it has been estimated that no more than 10\% of the amorphous
silicates (by mass) are pyroxene-glass. It is unlikely that this reflects a
genuine conversion of olivine to pyroxene glasses at the low temperatures of the
diffuse interstellar medium. Rather, we attribute this difference to an
incomplete sampling of the stellar sources that inject silicate dust into the
interstellar medium. In particular, supernovae are conceivably an important
source of interstellar dust. However, because of their scarcity in the solar
neighborhood, infrared spectra of SN dust ejecta are virtual non-existing. With
the launch of Spitzer and, particularly, its sensitive infrared spectrometer,
IRS, this situation will hopefully quickly be remedied. 

We note that, in our fits, interstellar amorphous silicates contain an
appreciable fraction of iron.  In contrast, crystalline silicates injected by
late-type stars are in the form of forsterite and enstatite -- the magnesium
rich end members of the olivine (Mg$_2$SiO$_4$) and pyroxene (MgSiO$_3$)
families.  Theoretically, the condensation behavior of iron in the ejecta of
late-type stars is somewhat different than for the Solar Nebula.  Indeed, at the
low pressures of AGB winds, iron may not condense out until after the magnesium
rich silicates have formed.  In that case, because of the high super-saturation
required for nucleation of metallic iron, rather than condensing as separate
iron-metal grains, iron may react with these silicates to form mixed
iron-magnesium silicates.  Because this reaction occurs below the glass
temperature of these silicates, this reaction would be expected to lead to the
formation of glassy silicates \nocite{tielens89miras} (Tielens 1989).  Within
this scheme, the crystalline silicates would form at higher temperatures, well
above the glass temperatures, by direct condensation from the gas phase.

As remarked above, the profile of the 9.7 \micron\ feature in dense clouds and
toward YSO's has additional absorption in the wing compared to that of the
diffuse ISM.   While the profiles  differ, the  amorphous silicates in these
different environments are generally assumed to be mineralogically similar. 
The additional absorption in dense clouds is generally attributed to a
non-volatile dust species, whereas for the YSO's the additional absorption is
accounted for by simple ice species. Such fits are however not unique. Some
processing of silicates may occur in the protoplanetary disks around
protostars, resulting in a different mineralogy. This material may then be
deposited into the surrounding molecular cloud through the powerful
protostellar winds and hence the general interstellar medium (Tielens 2003).
However, how much processing occurs in these environments and how well this
material is mixed into molecular clouds and the general interstellar medium is
presently not well understood. 

Recently, genuine silicate stardust grains have been isolated in interplanetary
dust particles(IDPs) \nocite{messenger_etal03} (Messenger \etal\ 2003).   Some of
these grains are crystalline while others are amorphous. While most presolar
grains found in this study of IDPs contain abundant O, Si, and Mg, their
mineralogy is currently unknown.  Only one grain is identified as forsterite. 
Moreover, some interstellar dust may have a very mundane isotopic (eg., Solar
abundances) composition and these grains would be difficult to separate from
Solar system condensates \nocite{tielens03} (Tielens 2003). Nevertheless, such
studies hold the promise of direct information on various relevant aspects of
interstellar silicates. We note that GEMS, Glass with Embedded Metals and
Sulphides, are an abundant component of anhydrous IDPs \nocite{bradley94}
(Bradley 1994). The detailed morphology and structure of these GEMS betrays an
extensive ion-irradiation history (Bradley 1994). While this may reflect long
exposure to the ISM, this irradiation may also have occurred in the early solar
system environment  \nocite{tielens03,westphal_bradley04} (Tielens 2003; Westphal
and Bradley 2004). The 9.7 \micron\ band in GEMS-rich thin sections of IDPs is
broad and featureless and, in that respect, resembles the 9.7 \micron\ features
observed towards YSOs, the Trapezium cluster, and $\mu$ Cep
\nocite{bradley_etal99} (Bradley \etal\ 1999). Of course, as demonstrated in this
paper, these interstellar 9.7 \micron\ features differ in detail. Nevertheless,
these initial laboratory studies of solar system materials, which may well be
directly related to interstellar silicate grains, are very promising.  It is
clear that many questions remain on the detailed structure and composition of
interstellar silicates.

\clearpage


\begin{deluxetable}{lcc}
\tablecaption{Observing parameters}
\tablewidth{0pt}
\tablehead{
\colhead{Source} & \colhead{TDT} & \colhead{AOT}
}
\startdata
WR 98A &  09401206 & AOT01 \\
WR 104 &  09901207 & AOT01 \\
WR 112 &  10201908 & AOT01 \\
WR 118 &  10802509 & AOT01 \\
\enddata
\end{deluxetable}




\begin{deluxetable}{lcc}
\tablecaption{Model continuum parameters for the WR stars}
\tablewidth{0pt}
\tablehead{
\colhead{Source} & \colhead{$T_o$} & \colhead{$T_1$} \\
\colhead{}       & \colhead{K}   & \colhead{K} 
}
\startdata
WR 98A & 790   & 180 \\
WR 104 & 840   & 110 \\
WR 112 & 680   & 90 \\
WR 118 & 1020  & 150\\
\enddata
\end{deluxetable}




\begin{deluxetable}{lcccc}
\tablecaption{Silicate Optical Depths and Visual Extinctions}
\tablewidth{0pt}
\tablehead{
\colhead{Source} & \colhead{$\tau_{9.7}$} & \colhead{$\tau_{18.5}$} &
\colhead{$\tau_{9.7}/\tau_{18.5}$} & \colhead{$A_V$}
}
\startdata
WR 98A	& 0.78	& 0.39	& 2.0 &	12.42  \\
WR 104	& 0.38	& 0.28	& 1.4 &	 6.50  \\
WR 112	& 0.63	& 0.32	& 2.0 &	11.03  \\
WR 118	& 0.78	& 0.45	& 1.7 &	11.20  
\enddata
\tablenotetext{a}{A(V) from van der Hucht 2001.}
\end{deluxetable}




\begin{deluxetable}{lcccc}
\tablecaption{Calculated Silicate Band Strengths (C$_{\rm ext}/V$ [$10^4$ cm$^{-1}$])}
\tablehead{
\colhead{Amorphous Silicate} & \colhead{Solid Spheres}  & \colhead{Solid CDE}  &
\colhead{Porous Spheres\tablenotemark{a}}    & \colhead{Porous CDE\tablenotemark{a}} 
}
\startdata
Olivine (MgFeSiO$_4$)                  & 0.880 & 0.926 & 0.616 & 0.620  \\
Pyroxene ((Mg$_{0.5}$Fe$_{0.5}$SiO$_3$) & 0.983 & 0.997 & 0.666 & 0.656 \\
\enddata
\tablenotetext{a}{Fifty percent porosity.}
\end{deluxetable}




\begin{deluxetable}{lcccc}
\tablecaption{Silicon Abundance Relative to Hydrogen ($\times10^{-5}$)}
\tablehead{
\colhead{Amorphous Silicate} & \colhead{Solid Spheres}  & \colhead{Solid CDE}  &
\colhead{Porous Spheres\tablenotemark{a}}    & \colhead{Porous CDE\tablenotemark{a}} 
}
\startdata
Olivine (MgFeSiO$_4$)       & 4.20               & 3.99               & 3.00                  & 2.98 \\
Pyroxene (Mg$_{0.5}$Fe$_{0.5}$SiO$_3$)    & 4.80  & 4.74               & 3.55                  & 3.60 \\
Olivine + Pyroxene   &  4.34     &  3.89 & 3.14 & 3.07 \\
                     & (3.28+1.06) & (1.52+2.37) & (0.87+2.27) & (0.27+2.80) \\
\enddata
\tablenotetext{a}{Fifty percent porosity.}
\end{deluxetable}




\clearpage
\begin{deluxetable}{lccccc}
\tablecaption{Limit on SiC in ISM}
\tablehead{
\colhead{Source} & \colhead{Olivine} & \colhead{Pyroxene} & \colhead{N(total
silicate)} & \colhead{N($\beta$-SiC)} &
\colhead{N(Si$_{\rm SiC}$)/N(Si$_{\rm silicates}$)} \\
 & \multicolumn{2}{c}{$\times10^{-5}$/H} & \multicolumn{2}{c}{$\times10^{16}$
 cm$^{-2}$} &
}
\startdata
WR 98a & 3.02 & 1.15 & 98.4 & 3.07 & 0.03 \\
WR 112 & 2.86 & 1.34 & 88.0 & 3.65 & 0.04 \\
\enddata
\end{deluxetable}




\begin{deluxetable}{ccc}
\tablecaption{Extinction curve for local ISM and Galactic
Center\tablenotemark{a}\tablenotemark{b}}
\tablewidth{0pt}
\tablehead{
\colhead{Wavelength} & \colhead{$A_{\lambda}/A_K$} & \colhead{$A_{\lambda}/A_K$}\\
\colhead{\micron}    & \colhead{local ISM} & \colhead{Galactic Center}
}
\startdata
 1.240 & 2.724 &  2.724   \\
 1.340 & 2.315 &  2.315   \\
 1.440 & 2.002 &  2.002   \\
 1.540 & 1.759 &  1.759   \\
 1.640 & 1.565 &  1.565  \\
 1.740 & 1.408 &  1.408   \\
 1.840 & 1.279 &  1.279   \\
 1.940 & 1.171 &  1.171   \\
 2.040 & 1.081 &  1.081   \\
 2.140 & 1.004 &  1.004
\enddata
\tablenotetext{a}{The complete version of this table is in the electronic edition of
the Journal.  The printed edition contains only a sample.}
\tablenotetext{b}{The extinction curves for the local ISM and GC are equal at
wavelengths less than 8 \micron.  Longward of 8 \micron, the extinction curves
for these environments differ; see \S4 and Fig.~3.}
\end{deluxetable}


\clearpage

\bibliographystyle{apj}
\bibliography{references}

\begin{thebibliography}{}

\bibitem[\protect\citeauthoryear{{Aitken} et~al.}{{Aitken}
  et~al.}{1980}]{aitken_etal80}
{Aitken}, D.~K., {Barlow}, M.~J., {Roche}, P.~F.,  \& {Spenser}, P.~M. 1980,
  \mnras, 192, 679

\bibitem[\protect\citeauthoryear{{Anders} \& {Zinner}}{{Anders} \&
  {Zinner}}{1993}]{anders_zinner93}
{Anders}, E.,  \& {Zinner}, E. 1993, Meteoritics, 28, 490

\bibitem[\protect\citeauthoryear{{Blanco} et~al.}{{Blanco}
  et~al.}{1994}]{blanco_etal94}
{Blanco}, A., {Borghesi}, A., {Fonti}, S.,  \& {Orofino}, V. 1994, \aap, 283,
  561

\bibitem[\protect\citeauthoryear{{Bohren} \& {Huffman}}{{Bohren} \&
  {Huffman}}{1983}]{bohren_huffman83}
{Bohren}, C.~F.,  \& {Huffman}, D.~R. 1983, Absorption and scattering of light
  by small particles (New York: Wiley)

\bibitem[\protect\citeauthoryear{{Bowey}, {Adamson}, \& {Whittet}}{{Bowey}
  et~al.}{1998}]{bowey_adamson_whittet98}
{Bowey}, J.~E., {Adamson}, A.~J.,  \& {Whittet}, D.~C.~B. 1998, \mnras, 298,
  131

\bibitem[\protect\citeauthoryear{{Bradley}}{{Bradley}}{1994}]{bradley94}
{Bradley}, J.~P. 1994, Science, 265, 925

\bibitem[\protect\citeauthoryear{{Bradley} et~al.}{{Bradley}
  et~al.}{1999}]{bradley_etal99}
{Bradley}, J.~P., et~al. 1999, Science, 285, 1716

\bibitem[\protect\citeauthoryear{{Bringa} et~al.}{{Bringa}
  et~al.}{2005}]{bringa_etal05}
{Bringa}, E., et~al. 2005, to be submitted to {\it ApJ}

\bibitem[\protect\citeauthoryear{{Bussoletti} et~al.}{{Bussoletti}
  et~al.}{1987}]{bussoletti_etal87}
{Bussoletti}, E., {Colangeli}, L., {Borghesi}, A.,  \& {Orofino}, V. 1987,
  \aaps, 70, 257

\bibitem[\protect\citeauthoryear{{Cami}}{{Cami}}{2002}]{cami02}
{Cami}, J. 2002, Ph.D.~Thesis

\bibitem[\protect\citeauthoryear{{Chiar} \& {Tielens}}{{Chiar} \&
  {Tielens}}{2001}]{chiar_tielens01}
{Chiar}, J.~E.,  \& {Tielens}, A. G. G.~M. 2001, \apjl, 550, L207

\bibitem[\protect\citeauthoryear{{Demyk} et~al.}{{Demyk}
  et~al.}{2000}]{demyk_etal00}
{Demyk}, K., {Dartois}, E., {Wiesemeyer}, H., {Jones}, A.~P.,  \&
  {d'Hendecourt}, L. 2000, \aap, 364, 170

\bibitem[\protect\citeauthoryear{{Demyk} et~al.}{{Demyk}
  et~al.}{1999}]{demyk_etal99}
{Demyk}, K., {Jones}, A.~P., {Dartois}, E., {Cox}, P.,  \& {d'Hendecourt}, L.
  1999, \aap, 349, 267

\bibitem[\protect\citeauthoryear{{Dorschner} et~al.}{{Dorschner}
  et~al.}{1995}]{dorschner_etal95}
{Dorschner}, J., {Begemann}, B., {Henning}, T., {Jaeger}, C.,  \& {Mutschke},
  H. 1995, \aap, 300, 503

\bibitem[\protect\citeauthoryear{{Figer}, {McLean}, \& {Morris}}{{Figer}
  et~al.}{1999}]{figer_mclean_morris99}
{Figer}, D.~F., {McLean}, I.~S.,  \& {Morris}, M. 1999, \apj, 514, 202

\bibitem[\protect\citeauthoryear{{Forrest}, {Gillett}, \& {Stein}}{{Forrest}
  et~al.}{1975}]{forrest_gillett_stein75}
{Forrest}, W.~J., {Gillett}, F.~C.,  \& {Stein}, W.~A. 1975, \apj, 195, 423

\bibitem[\protect\citeauthoryear{{Gillett} et~al.}{{Gillett}
  et~al.}{1975}]{gillett_etal75}
{Gillett}, F.~C., {Forrest}, W.~J., {Merrill}, K.~M., {Soifer}, B.~T.,  \&
  {Capps}, R.~W. 1975, \apj, 200, 609

\bibitem[\protect\citeauthoryear{{Groenewegen}}{{Groenewegen}}{1995}]{groenewe%
gen95}
{Groenewegen}, M.~A.~T. 1995, \aap, 293, 463

\bibitem[\protect\citeauthoryear{{Grossman} \& {Larimer}}{{Grossman} \&
  {Larimer}}{1974}]{grossman_larimer74}
{Grossman}, L.,  \& {Larimer}, J. 1974, Rev.\ Geophys.\ Space Phys, 12, 71

\bibitem[\protect\citeauthoryear{{Henning} et~al.}{{Henning}
  et~al.}{1995}]{henning_etal95}
{Henning}, T., {Begemann}, B., {Mutschke}, H.,  \& {Dorschner}, J. 1995, \aaps,
  112, 143

\bibitem[\protect\citeauthoryear{{Henning} et~al.}{{Henning}
  et~al.}{1999}]{henning_etal99}
{Henning}, T., {Il'In}, V.~B., {Krivova}, N.~A., {Michel}, B.,  \&
  {Voshchinnikov}, N.~V. 1999, \aaps, 136, 405

\bibitem[\protect\citeauthoryear{{Heras} \& {Hony}}{{Heras} \&
  {Hony}}{2005}]{heras_hony05}
{Heras}, A.~M.,  \& {Hony}, S. 2005, \aap, 439, 171

\bibitem[\protect\citeauthoryear{{Indebetouw} et~al.}{{Indebetouw}
  et~al.}{2005}]{indebetouw_etal05}
{Indebetouw}, R., et~al. 2005, \apj, 619, 931

\bibitem[\protect\citeauthoryear{{J\"ager} et~al.}{{J\"ager}
  et~al.}{1994}]{jaeger_etal94}
{J\"ager}, C., {Mutschke}, H., {Begemann}, B., {Dorschner}, J.,  \& {Henning},
  T. 1994, \aap, 292, 641

\bibitem[\protect\citeauthoryear{{Jones}, {Tielens}, \& {Hollenbach}}{{Jones}
  et~al.}{1996}]{jones_etal96}
{Jones}, A.~P., {Tielens}, A. G. G.~M.,  \& {Hollenbach}, D.~J. 1996, \apj,
  469, 740

\bibitem[\protect\citeauthoryear{{Jura}}{{Jura}}{1980}]{jura80}
{Jura}, M. 1980, \apj, 235, 63

\bibitem[\protect\citeauthoryear{{Kemper}}{{Kemper}}{2002}]{kemper02thesis}
{Kemper}, F. 2002, Ph.D. thesis, University of Amsterdam

\bibitem[\protect\citeauthoryear{{Kemper} et~al.}{{Kemper}
  et~al.}{2002}]{kemper_etal02}
{Kemper}, F., {de Koter}, A., {Waters}, L.~B.~F.~M., {Bouwman}, J.,  \&
  {Tielens}, A.~G.~G.~M. 2002, \aap, 384, 585

\bibitem[\protect\citeauthoryear{{Kemper}, {Vriend}, \& {Tielens}}{{Kemper}
  et~al.}{2004}]{kemper_vriend_tielens04}
{Kemper}, F., {Vriend}, W.~J.,  \& {Tielens}, A.~G.~G.~M. 2004, \apj, 609, 826

\bibitem[\protect\citeauthoryear{{Kemper}, {Vriend}, \& {Tielens}}{{Kemper}
  et~al.}{2005}]{kemper_vriend_tielens05}
{Kemper}, F., {Vriend}, W.~J.,  \& {Tielens}, A.~G.~G.~M. 2005, \apj, 609,
  submitted

\bibitem[\protect\citeauthoryear{{Leech} et~al.}{{Leech}
  et~al.}{2002}]{leech_etal02}
{Leech}, K., et~al. 2002, The ISO Handbook, Volume V: SWS - The Short
  Wavelength Spectrometer

\bibitem[\protect\citeauthoryear{{Malfait} et~al.}{{Malfait}
  et~al.}{1999}]{malfait_etal99}
{Malfait}, K., {Waelkens}, C., {Bouwman}, J., {de Koter}, A.,  \& {Waters},
  L.~B.~F.~M. 1999, \aap, 345, 181

\bibitem[\protect\citeauthoryear{{Mathis}}{{Mathis}}{1996}]{mathis96}
{Mathis}, J.~S. 1996, \apj, 472, 643

\bibitem[\protect\citeauthoryear{{Mathis}}{{Mathis}}{1998}]{mathis98}
{Mathis}, J.~S. 1998, \apj, 497, 824

\bibitem[\protect\citeauthoryear{{Mennella}, {Colangeli}, \&
  {Bussoletti}}{{Mennella} et~al.}{1995}]{mennella_etal95}
{Mennella}, V., {Colangeli}, L.,  \& {Bussoletti}, E. 1995, \aap, 295, 165

\bibitem[\protect\citeauthoryear{{Messenger} et~al.}{{Messenger}
  et~al.}{2003}]{messenger_etal03}
{Messenger}, S., {Keller}, L.~P., {Stadermann}, F.~J., {Walker}, R.~M.,  \&
  {Zinner}, E. 2003, Science, 300, 105

\bibitem[\protect\citeauthoryear{{Moneti} et~al.}{{Moneti}
  et~al.}{2001}]{moneti_etal01}
{Moneti}, A., {Stolovy}, S., {Blommaert}, J.~A.~D.~L., {Figer}, D.~F.,  \&
  {Najarro}, F. 2001, \aap, 366, 106

\bibitem[\protect\citeauthoryear{{Ossenkopf}}{{Ossenkopf}}{1991}]{ossenkopf91}
{Ossenkopf}, V. 1991, \aap, 251, 210

\bibitem[\protect\citeauthoryear{{Posch} et~al.}{{Posch}
  et~al.}{2002}]{posch_etal02}
{Posch}, T., {Kerschbaum}, F., {Mutschke}, H., {Dorschner}, J.,  \&
  {J{\"a}ger}, C. 2002, \aap, 393, L7

\bibitem[\protect\citeauthoryear{{Roche} \& {Aitken}}{{Roche} \&
  {Aitken}}{1984}]{roche_aitken84}
{Roche}, P.~F.,  \& {Aitken}, D.~K. 1984, \mnras, 208, 481

\bibitem[\protect\citeauthoryear{{Roche} \& {Aitken}}{{Roche} \&
  {Aitken}}{1985}]{roche_aitken85}
{Roche}, P.~F.,  \& {Aitken}, D.~K. 1985, \mnras, 215, 425

\bibitem[\protect\citeauthoryear{{Salpeter}}{{Salpeter}}{1977}]{salpeter77}
{Salpeter}, E.~E. 1977, \araa, 15, 267

\bibitem[\protect\citeauthoryear{{Sembach} \& {Savage}}{{Sembach} \&
  {Savage}}{1996}]{sembach_savage96}
{Sembach}, K.~R.,  \& {Savage}, B.~D. 1996, \apj, 457, 211

\bibitem[\protect\citeauthoryear{{Smith}, {Aitken}, \& {Roche}}{{Smith}
  et~al.}{1990}]{smith_aitken_roche90}
{Smith}, C.~H., {Aitken}, D.~K.,  \& {Roche}, P.~F. 1990, \mnras, 246, 1

\bibitem[\protect\citeauthoryear{{Smith} et~al.}{{Smith}
  et~al.}{2000}]{smith_etal00}
{Smith}, C.~H., {Wright}, C.~M., {Aitken}, D.~K., {Roche}, P.~F.,  \& {Hough},
  J.~H. 2000, \mnras, 312, 327

\bibitem[\protect\citeauthoryear{{Snow} \& {Witt}}{{Snow} \&
  {Witt}}{1996}]{snow_witt96}
{Snow}, T.~P.,  \& {Witt}, A.~N. 1996, \apjl, 468, L65

\bibitem[\protect\citeauthoryear{{Sofia} \& {Meyer}}{{Sofia} \&
  {Meyer}}{2001}]{sofia_meyer01}
{Sofia}, U.~J.,  \& {Meyer}, D.~M. 2001, \apjl, 554, L221

\bibitem[\protect\citeauthoryear{{Speck}, {Barlow}, \& {Skinner}}{{Speck}
  et~al.}{1997}]{speck_barlow_skinner97}
{Speck}, A.~K., {Barlow}, M.~J.,  \& {Skinner}, C.~J. 1997, \mnras, 288, 431

\bibitem[\protect\citeauthoryear{{Sylvester} et~al.}{{Sylvester}
  et~al.}{1999}]{sylvester_etal99}
{Sylvester}, R.~J., {Kemper}, F., {Barlow}, M.~J., {de Jong}, T., {Waters},
  L.~B.~F.~M., {Tielens}, A.~G.~G.~M.,  \& {Omont}, A. 1999, \aap, 352, 587

\bibitem[\protect\citeauthoryear{{Tielens}}{{Tielens}}{1989}]{tielens89miras}
{Tielens}, A.~G.~G.~M. 1989, in From Miras to planetary nebulae: Which path for
  stellar evolution (Paris: Edition Frontieres), 186

\bibitem[\protect\citeauthoryear{{Tielens}}{{Tielens}}{1990}]{tielens90}
{Tielens}, A.~G.~G.~M. 1990, in Carbon in the Galaxy: Studies from Earth and
  Space, 59

\bibitem[\protect\citeauthoryear{{Tielens}}{{Tielens}}{1997}]{tielens97}
{Tielens}, A.~G.~G.~M. 1997, \apss, 251, 1

\bibitem[\protect\citeauthoryear{{Tielens}}{{Tielens}}{2003}]{tielens03}
{Tielens}, A.~G.~G.~M. 2003, Science, 300, 68

\bibitem[\protect\citeauthoryear{{Tielens} \& {Allamandola}}{{Tielens} \&
  {Allamandola}}{1987}]{tielens_allamandola87}
{Tielens}, A.~G.~G.~M.,  \& {Allamandola}, L.~J. 1987, in Physical processes in
  interstellar clouds; Proceedings of the NATO Advanced Study Institute, Irsee,
  Federal Republic of Germany, Aug. 18-28, 1986 (Dordrecht: D. Reidel
  Publishing Co.), 333

\bibitem[\protect\citeauthoryear{{Tsuji}}{{Tsuji}}{2000}]{tsuji00}
{Tsuji}, T. 2000, \apjl, 540, L99

\bibitem[\protect\citeauthoryear{{van Boekel} et~al.}{{van Boekel}
  et~al.}{2004}]{vanboekel_etal04}
{van Boekel}, R., et~al. 2004, \nat, 432, 479

\bibitem[\protect\citeauthoryear{{van der Hucht} et~al.}{{van der Hucht}
  et~al.}{1996}]{vanderhucht_etal96}
{van der Hucht}, K.~A., et~al. 1996, \aap, 315, L193

\bibitem[\protect\citeauthoryear{{Waters} \& {Waelkens}}{{Waters} \&
  {Waelkens}}{1998}]{waters_waelkens98}
{Waters}, L.~B.~F.~M.,  \& {Waelkens}, C. 1998, \araa, 36, 233

\bibitem[\protect\citeauthoryear{{Westphal} \& {Bradley}}{{Westphal} \&
  {Bradley}}{2004}]{westphal_bradley04}
{Westphal}, A.~J.,  \& {Bradley}, J.~P. 2004, \apj, 617, 1131

\bibitem[\protect\citeauthoryear{{Whittet}}{{Whittet}}{2003}]{whittet_bk03}
{Whittet}, D. C.~B. 2003, Dust in the Galactic Environment (2nd ed.) (Bristol:
  Institute of Physics (IOP) Publishing)

\bibitem[\protect\citeauthoryear{{Whittet} et~al.}{{Whittet}
  et~al.}{1988}]{whittet_etal88}
{Whittet}, D. C.~B., {Bode}, M.~F., {Longmore}, A.~J., {Admason}, A.~J.,
  {McFadzean}, A.~D., {Aitken}, D.~K.,  \& {Roche}, P.~F. 1988, \mnras, 233,
  321

\bibitem[\protect\citeauthoryear{{Whittet}, {Duley}, \& {Martin}}{{Whittet}
  et~al.}{1990}]{whittet_duley_martin90}
{Whittet}, D.~C.~B., {Duley}, W.~W.,  \& {Martin}, P.~G. 1990, \mnras, 244, 427

\bibitem[\protect\citeauthoryear{{Williams}, {van der Hucht}, \&
  {Th\'e}}{{Williams} et~al.}{1987}]{williams_vanderhucht_the87}
{Williams}, P., {van der Hucht}, K.,  \& {Th\'e}, P. 1987, \aap, 182, 91

\bibitem[\protect\citeauthoryear{{Wolff}, {Clayton}, \& {Meade}}{{Wolff}
  et~al.}{1993}]{wolff_clayton_meade93}
{Wolff}, M.~J., {Clayton}, G.~C.,  \& {Meade}, M.~R. 1993, \apj, 403, 722

\end{thebibliography}

\clearpage
\begin{figure}
\epsscale{0.7}
\plotone{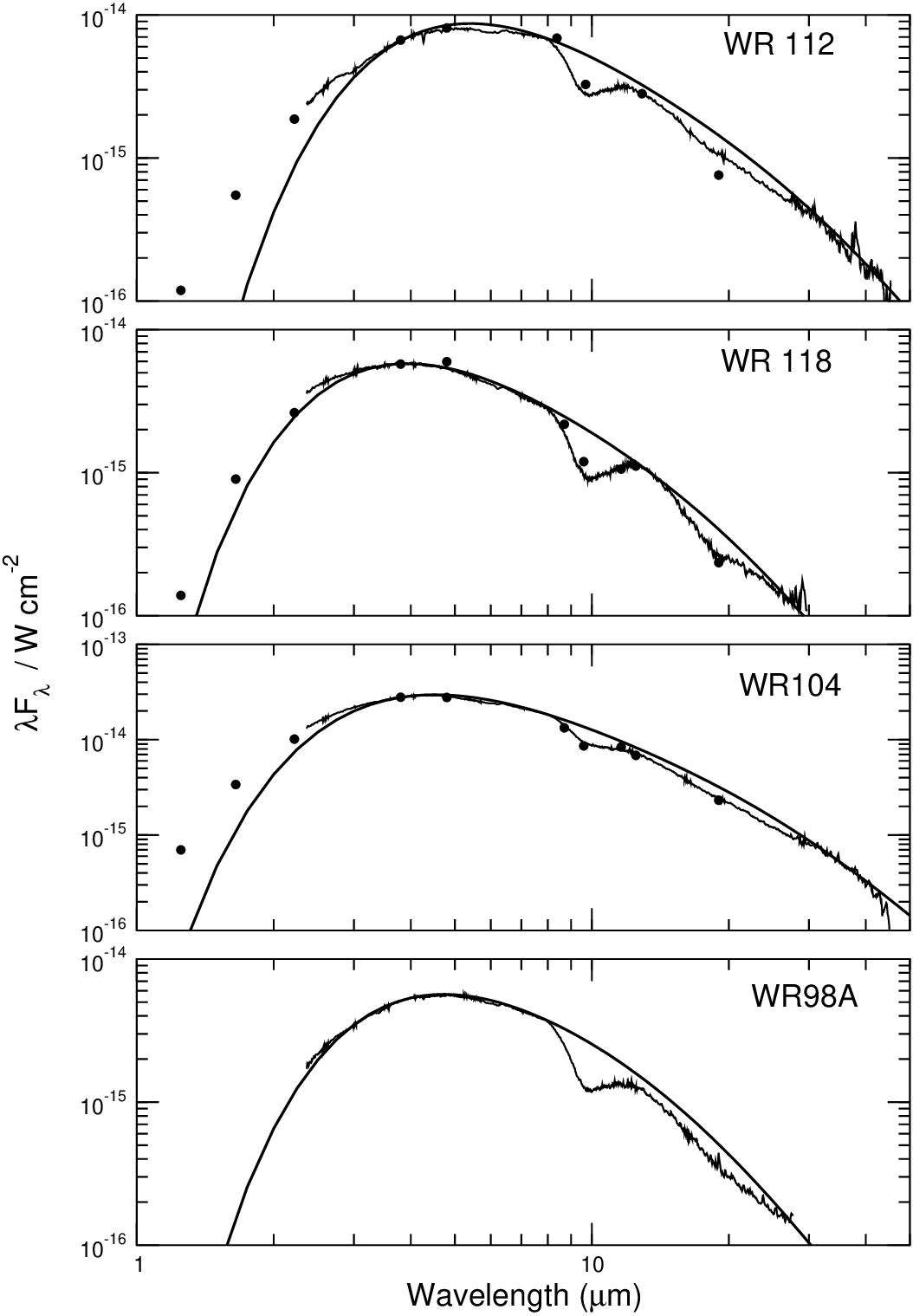}
\caption{2.5-45 \micron\ ISO-SWS AOT1 spectra for heavily extincted WR stars WR
112, WR 118, WR 104, and WR 98A.  All spectra show the 9.7 and 18 \micron\
silicate absorption.  Also shown are photometry points from Williams \etal\
1987 (where available), normalized to the ISO spectrum in the $L$ band.  The
continuum model is described in the text. \label{fig:model}}
\end{figure}

\begin{figure}
\epsscale{0.8}
\plotone{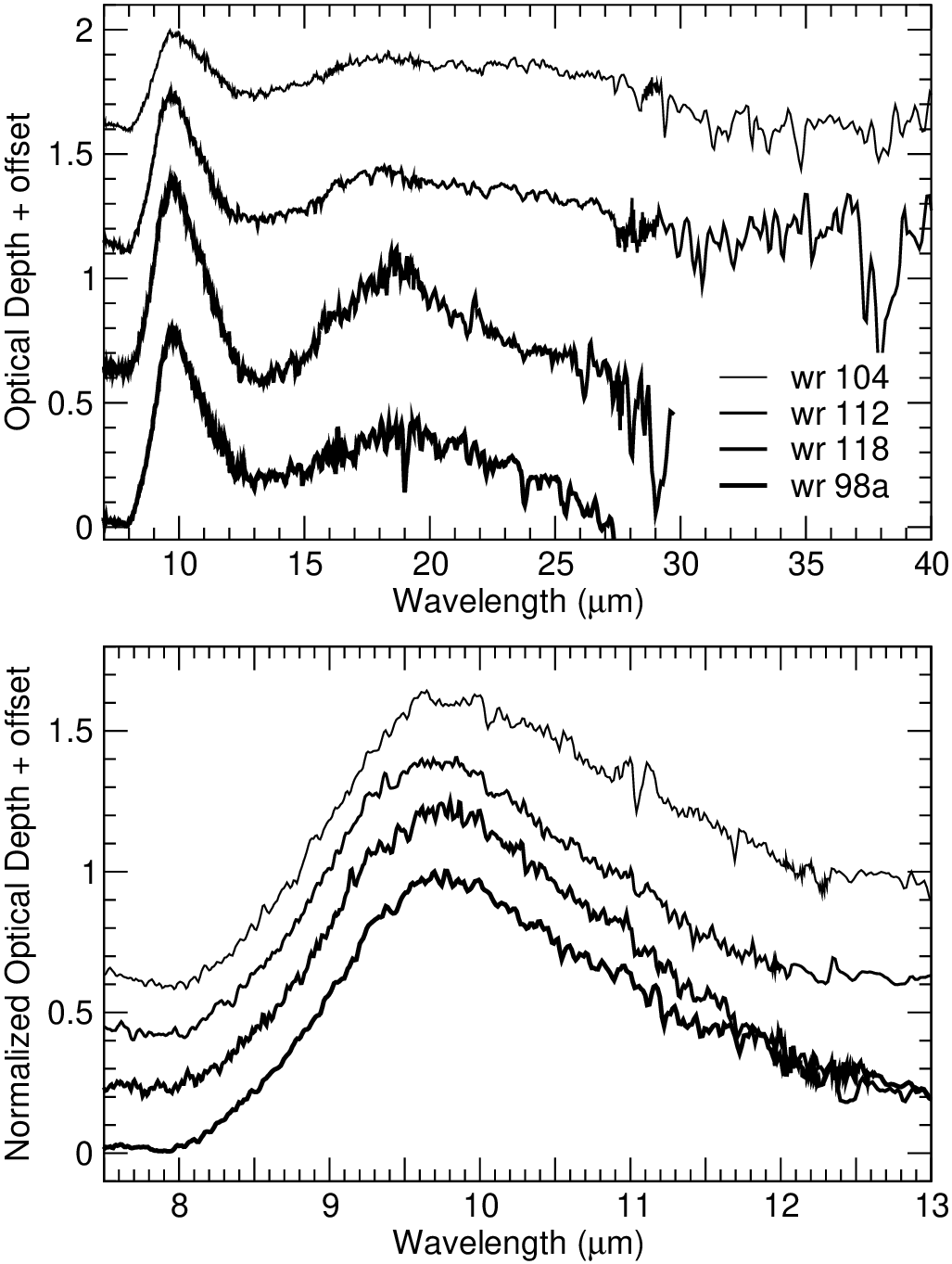}
\caption{Optical depth absorption spectra resulting from model continuum fits
shown in Figure 1.  The order of spectra from top to bottom is WR 104, WR 112,
WR 118 and WR 98a in both panels. [Top] The 7 to 40 \micron\ region. The
spectra are offset from each other for clarity. [Bottom] The 7.5 to 13 \micron\
region of the silicate feature.  The  spectra have been normalized to
$\tau_{9.7}=1.0$ and each spectrum is offset by 0.2 optical depth units from
the spectrum below.  \label{fig:taus}}
\end{figure}

\begin{figure}[hbt]
\epsscale{1.0}
\plotone{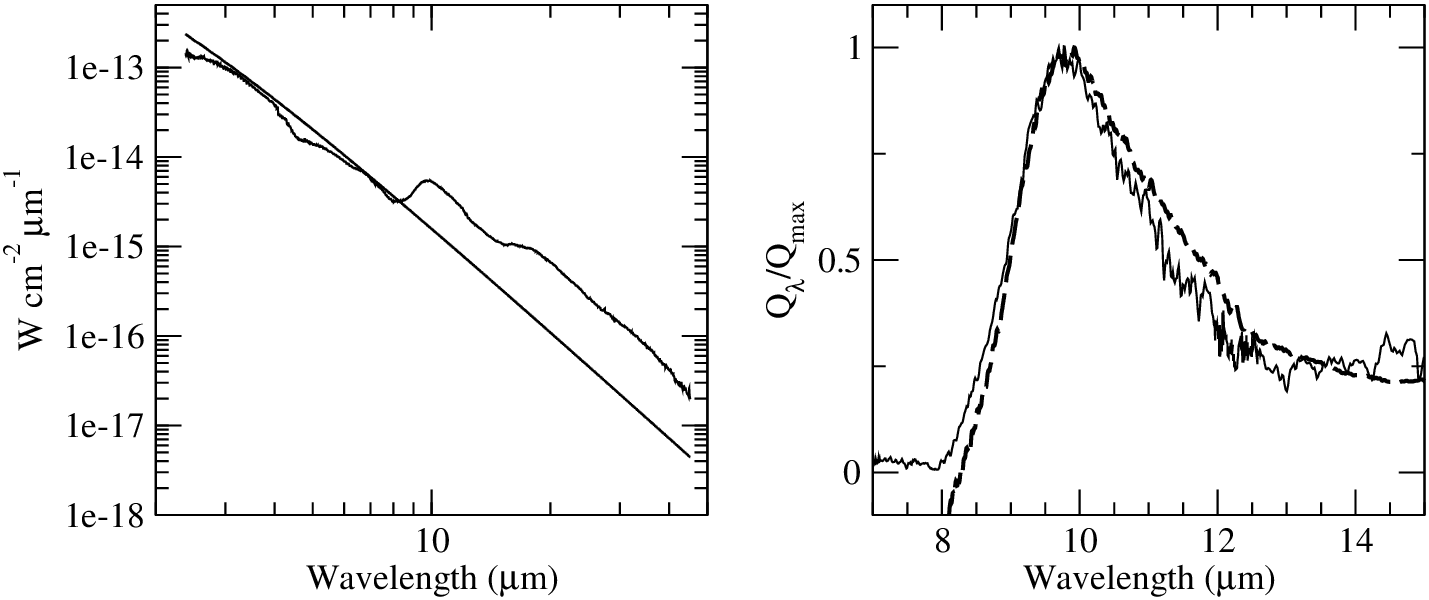}
\caption{[left] The dereddened ($A_V = 1.5$) ISO-SWS spectrum of $\mu$ Cep
shown with a 3600 K blackbody representing the stellar continuum. [right,
dashed line] The $\mu$ Cep emissivity curve  derived from the spectrum at
the left assuming $T_d = 250$ K for the underlying dust emission, compared with
the silicate absorption feature of WR 98a [solid line].
\label{fig:mucep}}
\end{figure}

\begin{figure}[hbt]
\epsscale{.9}
\plotone{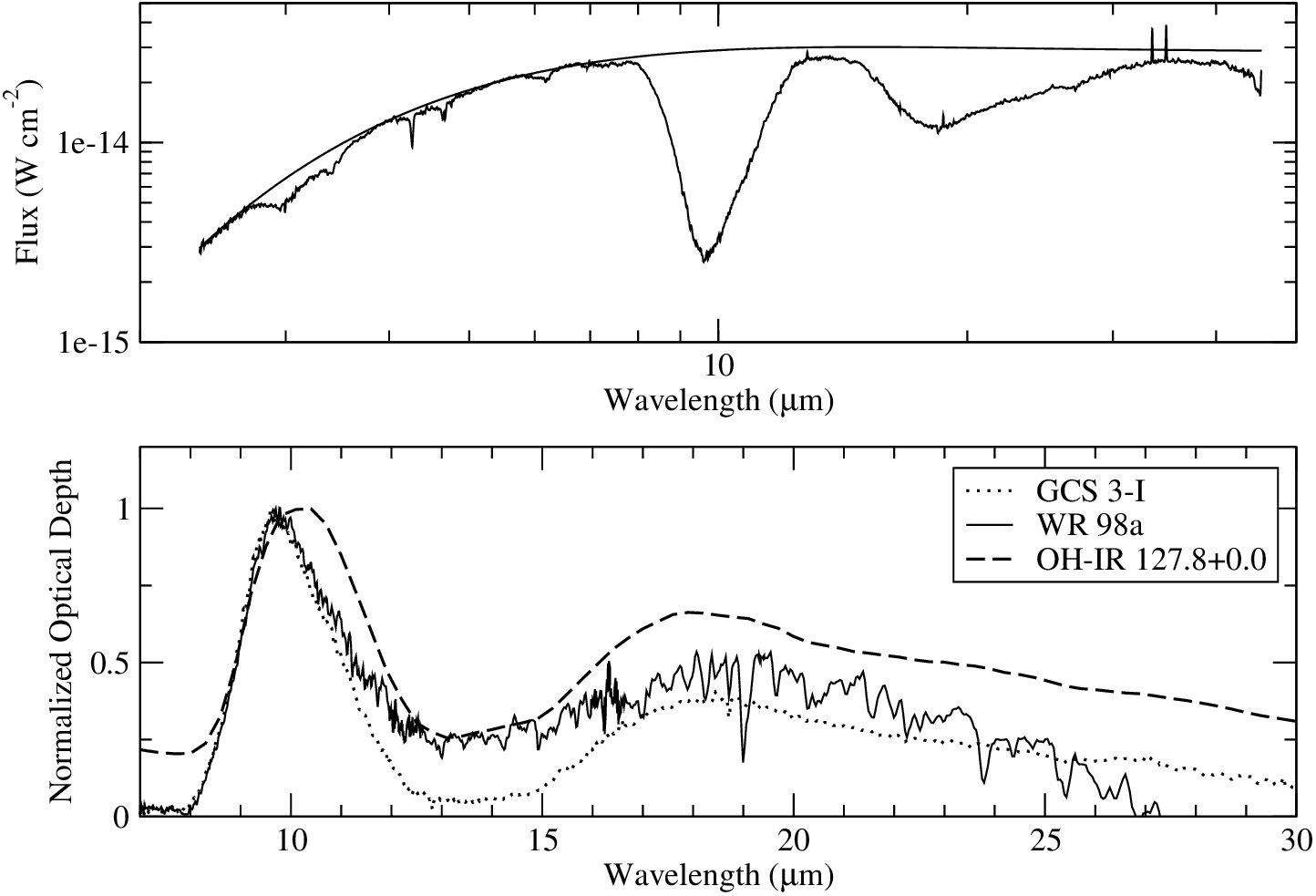}
\caption{[Top] ISO-SWS spectrum of GCS 3-I with fourth order polynomial
continuum used to derive the optical depth spectrum in the panel below. [Bottom]
Extinction model profile for OH-IR 127.8+0.0 (dashed line; Kemper et al.
2002).  Extinction profile for the Galactic Center Quintuplet source GCS 3-I
(dotted line).  Extinction profile for WR 98a (solid line).
\label{fig:ohir_gcs3}}
\end{figure}

\begin{figure}[hbt]
\epsscale{1}
\plotone{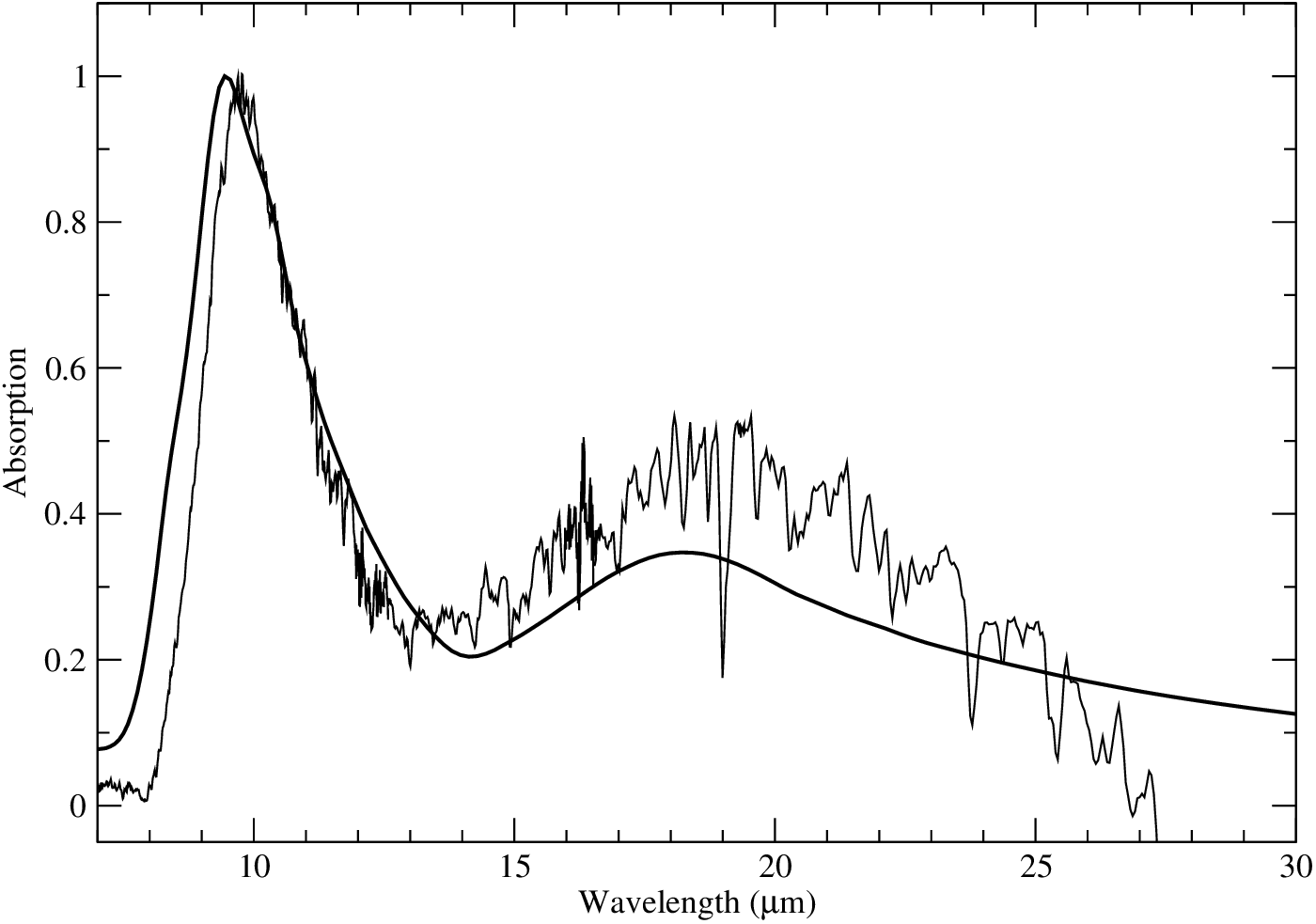}
\caption{The extinction profiles for Drain \& Lee ``astronomical silicates''
compared to the WR 98a absorption profile.
\label{fig:draine}}
\end{figure}

\begin{figure}
\epsscale{1.0}
\plotone{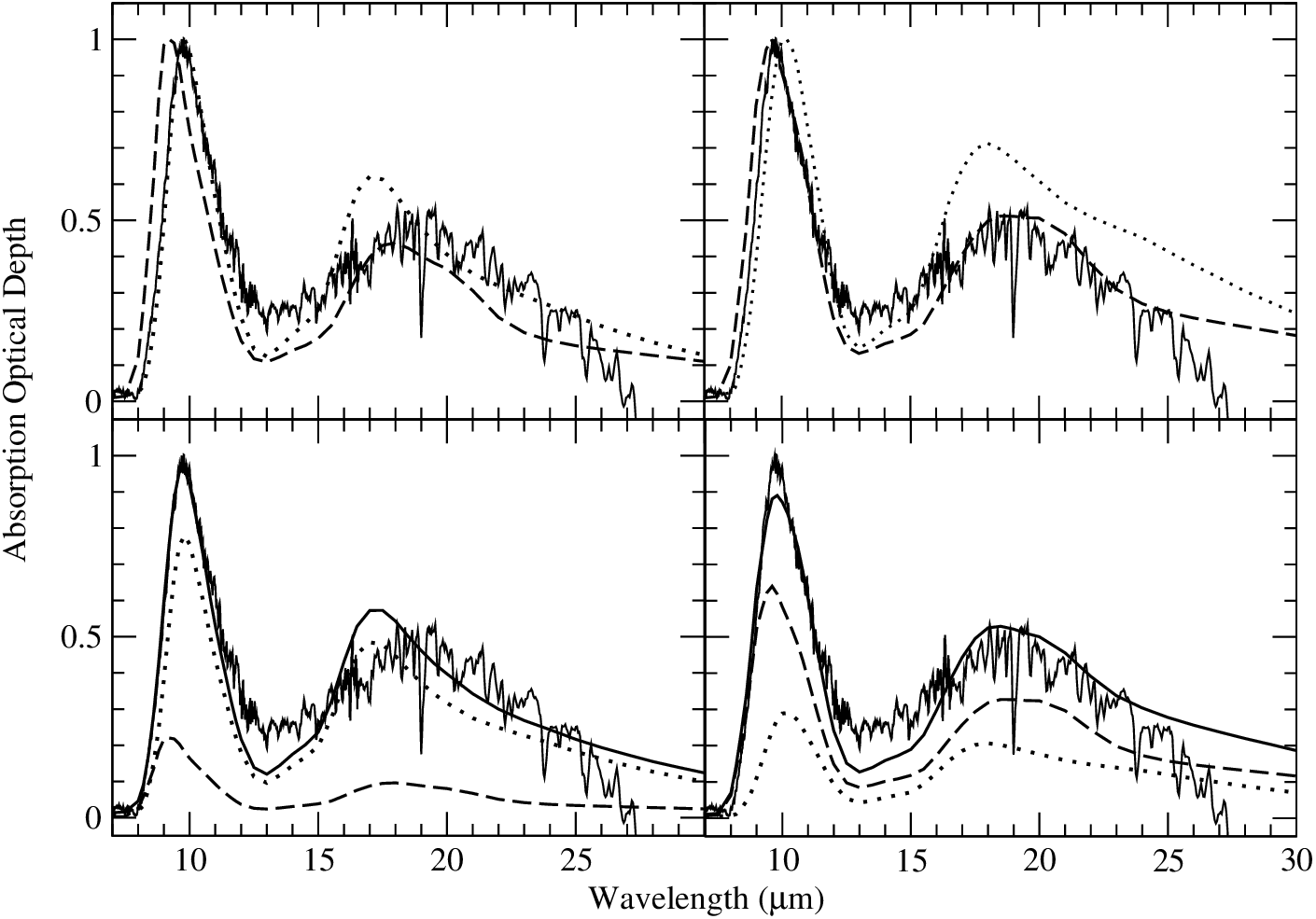}
\caption{The absorption profile for WR 98a normalized to unity at 9.7 \micron\
compared with the extinction profiles for amorphous olivine-glass (dotted line) and
pyroxene-glass (dashed line) spheres.  Solid [left panels] and porous [right panels]
spheres are considered.   The porous spheres contain 50\% vacuum.  The bottom
panels show the effect of combining separate populations of olivine-glass and pyroxene-glass
spherical grains; the solid smooth line represents the total absorption profile.
\label{fig:spheres}}
\end{figure}

\begin{figure}
\epsscale{1.0}
\plotone{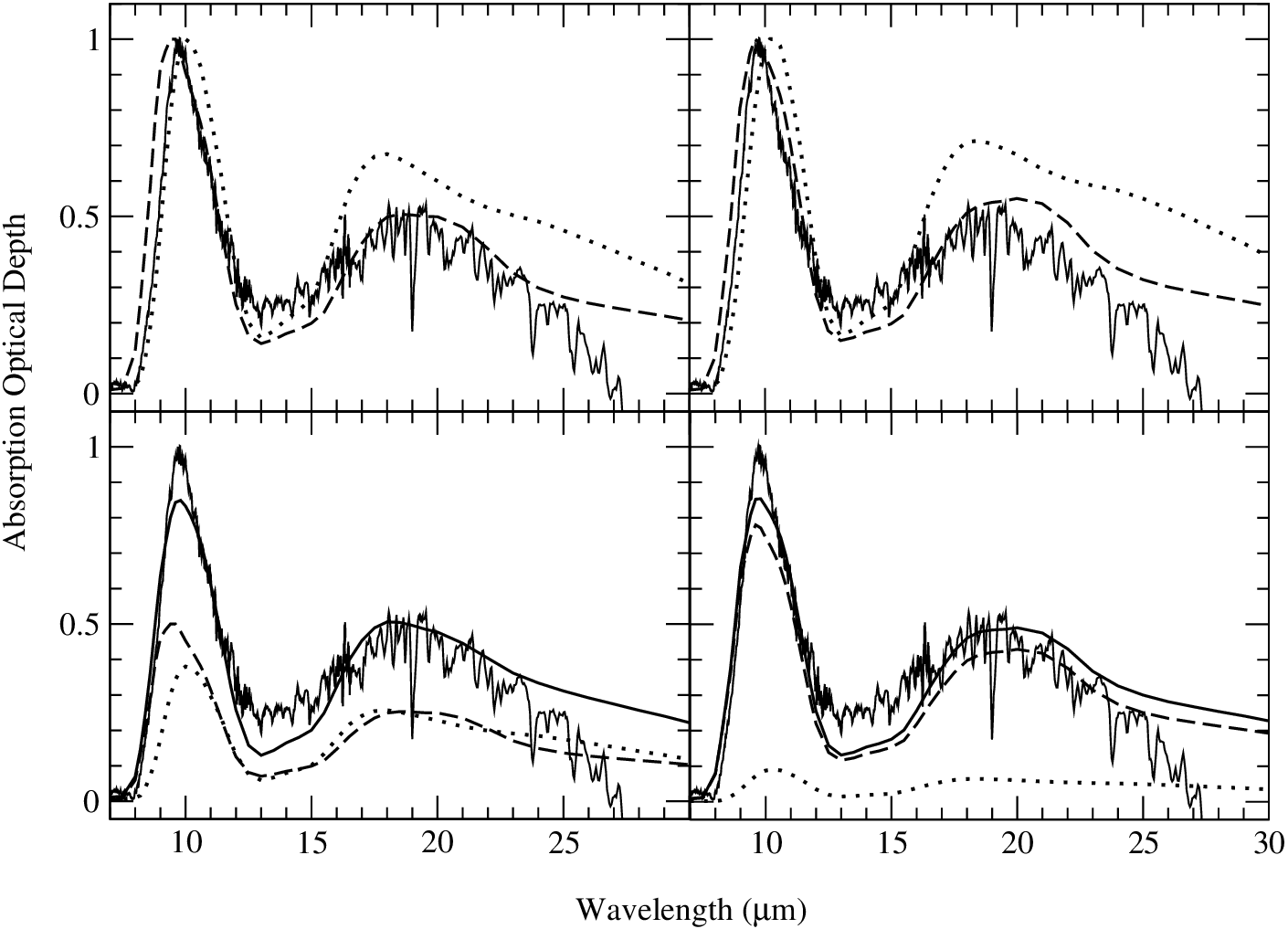}
\caption{The absorption profile for WR 98a normalized to unity at 9.7 \micron\
compared with the extinction profiles for amorphous olivine-glass (dotted line) and
pyroxene-glass (dashed line) CDE.  Solid [left panels] and porous [right panels]
CDE are considered.   The porous CDE contain 50\% vacuum.  The bottom
panels show the effect of combining separate populations of olivine-glass and pyroxene-glass
CDE; the solid smooth line represents the total absorption profile.
\label{fig:cde}}
\end{figure}

\begin{figure}[hbt]
\epsscale{0.8}
\plotone{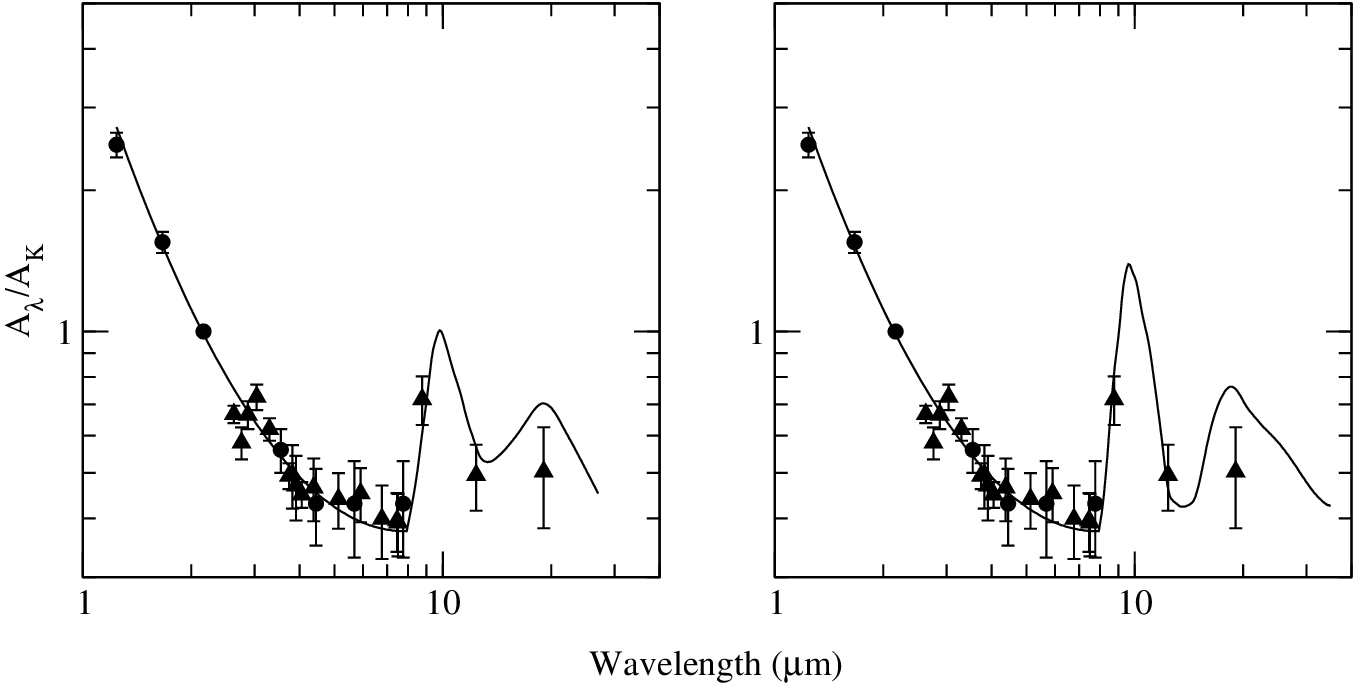}
\caption{Interstellar extinction curves for the local ISM (left) and the
Galactic Center (right).  Circles represent measured extinction for the
Galactic plane (Indebetouw \etal\ 2005.  Triangles represent the extinction
measured toward the Galactic Center (Lutz 1999).  The smooth line from 1.25 to
8 \micron\ is a least-squares fit to the measured extinction points.  The
silicate curves are representations of the silicate features for the local ISM
(WR 98a; left) and the Galactic Center (GCS3; right).  \label{fig:extinction}}
\end{figure}

\end{document}